\documentclass[twocolumn]{aastex631}
\shorttitle{The central region of NGC~4395} 
\shortauthors{Payel Nandi et al.}

\graphicspath{{./}{figures/}}

\begin{document}

\title{Evidence for low power radio jet$-$ISM interaction at 10 parsec in the dwarf AGN host NGC~4395}

\author{Payel Nandi}{0009-0003-9765-3517}
\affiliation{Joint Astronomy Programme, Department of Physics, Indian Institute of Science, Bangalore 560012, India}
\affiliation{Indian Institute of Astrophysics, Block II, Koramangala, Bangalore 560034, India}
\author{C. S. Stalin}
\affiliation{Indian Institute of Astrophysics, Block II, Koramangala, Bangalore 560034, India}
\author{D. J. Saikia}
\affiliation{Inter-University Centre for Astronomy and Astrophysics, Pune 411007, India}
\author{Rogemar A. Riffel}
\affiliation{Departamento de Física, CCNE, Universidade Federal de Santa Maria, 97105-900, Santa Maria, RS, Brazil}
\author{Arijit Manna}
\affiliation{Midnapore City College, Kuturia, Bhadutala, Paschim Medinipur, West Bengal, 721129, India}
\author{Sabyasachi Pal}
\affiliation{Midnapore City College, Kuturia, Bhadutala, Paschim Medinipur, West Bengal, 721129, India}
\author{O. L. Dors}
\affiliation{UNIVAP - Universidade do Vale do Paraíba. Av. Shishima Hifumi, 2911, CEP: 12244-000 São José dos Campos, SP, Brazil}
\author{Dominika Wylezalek}
\affiliation{Astronomisches Rechen-Institut, Zentrum fur Astronomie der Universitat Heidelberg, Monchhofstr. 12-14, 69120 Heidelberg, Germany}
\author{Vaidehi S. Paliya}
\affiliation{Inter-University Centre for Astronomy and Astrophysics, Pune 411007, India}
\author{ P. Saikia}
\affiliation{Center for Astro, Particle and Planetary Physics, New York University Abu Dhabi, PO Box 129188, Abu Dhabi, UAE}
\author{Pratik Dabhade}
\affiliation{Instituto de Astrofísica de Canarias, Calle Vía Láctea, s/n, E-38205, La Laguna, Tenerife, Spain}
\author{Markus-Kissler Patig}
\affiliation{ESA - ESAC - European Space Agency, Camino Bajo del Castillo s/n, 28692 Villafranca del Castillo, Madrid, Spain}
\author{Ram Sagar}
\affiliation{Indian Institute of Astrophysics, Block II, Koramangala, Bangalore 560034, India}

\correspondingauthor{Payel Nandi}
\email{payel.nandi@iiap.res.in}

\begin{abstract}
Black hole driven outflows in galaxies hosting active galactic nuclei (AGN) 
may interact with their interstellar medium (ISM) affecting star formation. 
Such feedback processes, reminiscent of those seen in massive galaxies, have 
been reported recently in some dwarf galaxies. However, such studies have usually 
been on kiloparsec and larger scales and our knowledge on the smallest spatial
scales to which these feedback processes can operate is unclear. 
Here we demonstrate radio jet$-$ISM interaction on the scale of an 
asymmetric triple radio structure of $\sim$ 10 parsec size in NGC~4395. This triple radio structure
is seen in the 15 GHz continuum image and the two asymmetric jet-like structures
are situated on either side of the radio core that coincides with the optical
{\it Gaia} position.  The high resolution radio image and 
the extended [OIII]$\lambda$5007 emission, indicative of an outflow, are 
spatially coincident and are consistent with the interpretation of a low
power radio jet interacting with the ISM. Modelling of the spectral lines using {\tt MAPPINGS}, and estimation of temperature using optical integral
field spectroscopic data suggest shock ionization of the gas. The continuum 
emission at 237 GHz, though weak, was found to spatially coincide with the AGN. However, the CO(2$-$1) line emission was found 
to be displaced by around 20 parsec northward of the AGN core.  The spatial 
coincidence of molecular H$_2$$\lambda$2.4085 along the jet direction, the morphology of 
ionised [OIII]$\lambda$5007 and displacement of the CO(2$-$1) emission argues 
for conditions less favourable for star formation in the central $\sim$ 10 parsec region.
\end{abstract}
\keywords{Dwarf galaxies (416) --- Active galactic nuclei (16) --- radio jets (1347) --- AGN 
host galaxies (2017)}

\section{Introduction} \label{sec:intro}

\noindent
Active galactic nuclei (AGN), powered by the accretion of matter onto supermassive black holes at the centers of galaxies \citep{1984ARA&A..22..471R}, affect their host galaxies through the feedback processes.  They can have an impact on the interstellar medium (ISM) of their hosts via energetic outflows \citep{2021A&A...648A..17V}. 
These outflows, driven by radiation pressure, jets or winds from AGN, can occur from accretion disk to galaxy scales \citep{2018MNRAS.479.5544M,2021A&A...656A..55M}, 
and  can inhibit (negative feedback; \citealt{2012MNRAS.425L..66M}) or enhance star formation (SF) (positive feedback; \citealt{2020A&A...639L..13N}, \citealt{2017Natur.544..202M}). Both positive and negative feedback processes are also seen in a single system  \citep{2019ApJ...881..147S,2021A&A...645A..21G}.
Outflows affecting SF through the interaction of radio jets with the ISM in the host galaxies are known for large massive galaxies \citep[for reviews, see][]{2012ARA&A..50..455F,2022JApA...43...97S}. Recent observational evidence of dwarf galaxies hosting an AGN \citep{2022Natur.601..329S} challenges theoretical models that generally invoke supernovae feedback in dwarf galaxies \citep{2022MNRAS.516.2112K, 2023ApJ...950...81N}. Also, recently, outflows have been observed in an AGN hosted by dwarf galaxies \citep{2019ApJ...884...54M,2021ApJ...911...70B}.
Available observations of AGN have usually identified the impact of jets on the ISM on kpc or larger scales.  To have a clear understanding of the effect of jets on the ISM, one needs to study their impact from sub-pc to kpc and larger scales.  
There is hardly any observational evidence of jet$-$ISM interaction and its 
impact on the host galaxies of AGN on parsec scales.

NGC~4395 is a bulgeless dwarf galaxy at a distance of 4.3$\pm$0.3 Mpc 
\citep{2004AJ....127.2322T} and hosting a radio-quiet  AGN \citep{1989ApJ...342L..11F}. Its nucleus has the optical spectrum characteristic of a Seyfert 1.8 type AGN \citep{2006A&A...455..773V}, hosts a 
black hole \citep{1993ApJ...410L..75F} in the mass range of 
10$^4$$-$10$^5$ M$_{\odot}$ \citep{2005ApJ...632..799P,2019NatAs...3..755W}, is   variable in X-ray wavelengths \citep{2005MNRAS.356..524V}. 
NGC~4395 appeared unresolved in the Very Large Array (VLA) 
A-configuration image at 
1.4 GHz with a flux density of 1.68 mJy \citep{2001ApJS..133...77H}. 
In this paper, we present the first observational
evidence of a low power  radio jet in NGC~4395 interacting with 
the ISM of its host galaxy, driving shocks and  ionised outflow on 
parsec scale. Adopting a cosmology of H$_0$ = 70 km s$^{-1}$ Mpc$^{-1}$, 
$\Omega_M$ = 0.7,  $\Omega_{vac}$ = 0.3 and a distance of 4.3$\pm$0.3 Mpc
\citep{2004AJ....127.2322T}, 1$^{\prime\prime}$ in NGC~4395 corresponds 
to 21 parsec.

\begin{figure}
    \centering
\hspace*{-1.0cm}    \includegraphics[scale=0.5 ]{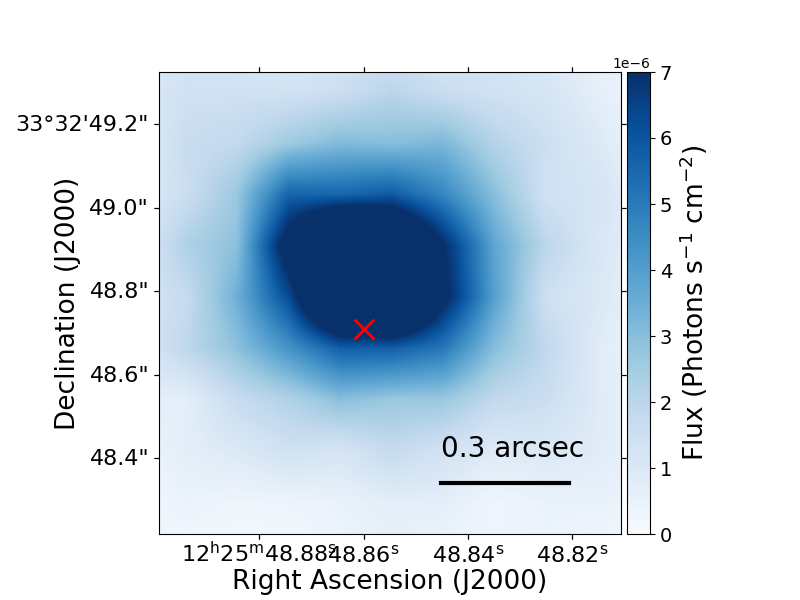}
    \caption{X-ray Image of NGC~4395 in the 0.5$-$7 keV energy range from {\it Chandra}. The red cross is the optical {\it Gaia} position.}
    \label{figure-1}
\end{figure}

\begin{figure*}
\includegraphics[scale=0.26]{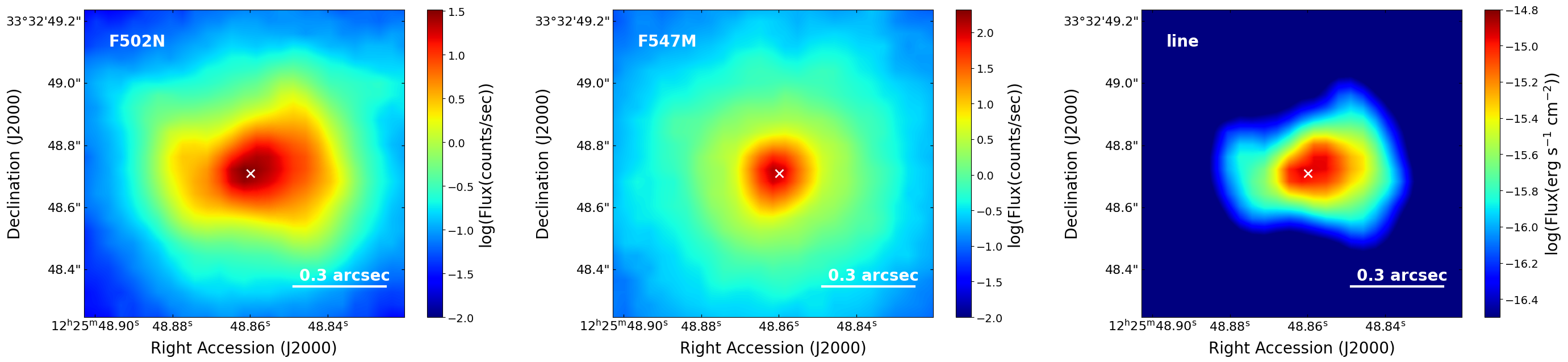}
\caption{Left panel: The narrow-band image from HST in the F502N filter that 
contains the emission from both the ionised [OIII]$\lambda$5007  gas and the 
continuum. Middle panel: The continuum image from HST in the F547M filter. Right 
panel: Difference image after subtraction of the scaled continuum F547M filter 
image from the narrow-band F502N filter image. This difference image reveals 
an asymmetric biconical [OIII]$\lambda$5007 outflow. The white cross in all the figures} is the optical core ({\it Gaia} position).

\label{figure-3}
\end{figure*}

\begin{figure*}
    \centering
    \hbox{
    \includegraphics[scale=0.38 ]{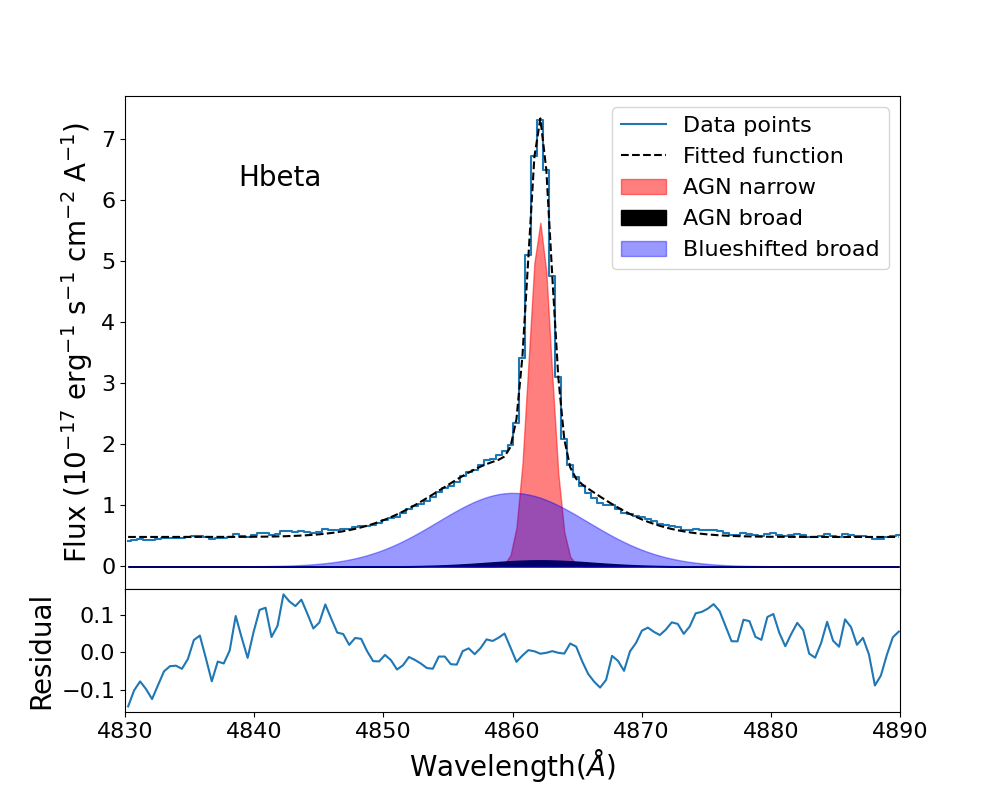}
    \hspace{-0.8 cm} \includegraphics[scale=0.38 ]{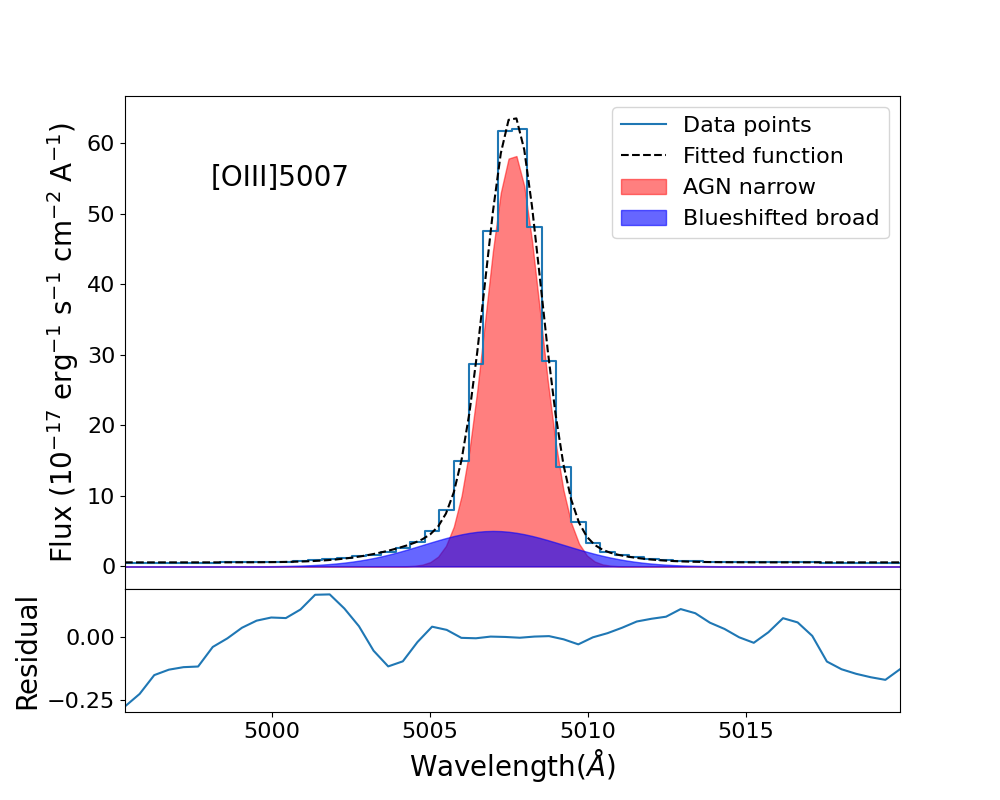}
    }
    \caption{Gaussian fit to the observed H$\beta$ (left) and [OIII]$\lambda$5007 (right) emission lines along
with the residuals (lower panel of each spectrum). Here, shaded blue is the Gaussian fit to the outflowing component, red and black are the Gaussian fits to the narrow and broad components, and the dotted black line
is the best fit spectrum. The observed spectra are shown as a solid blue line.}
    \label{figure-4}
\end{figure*}

\begin{figure*}
    \centering
    \includegraphics[scale=0.3]{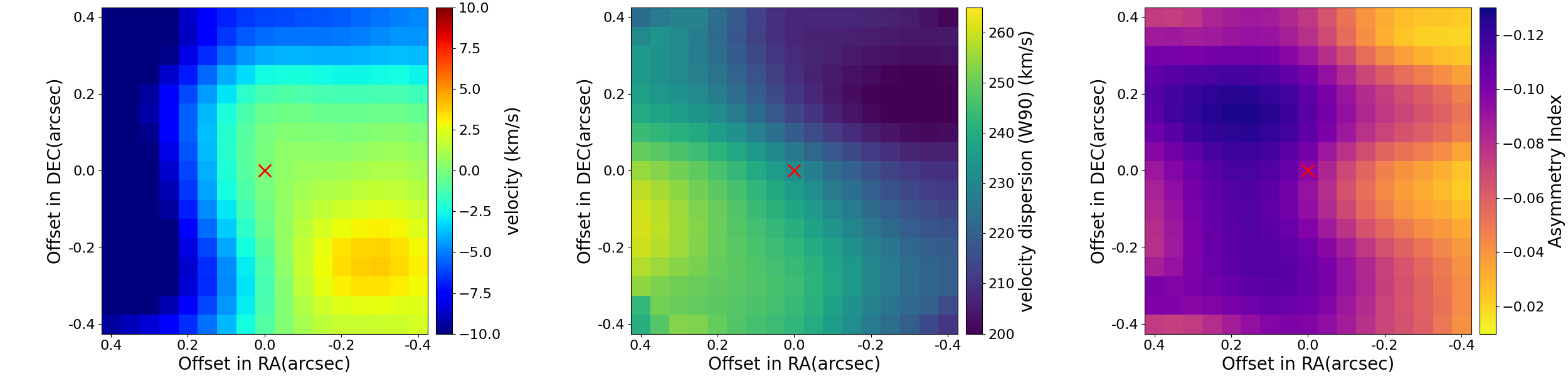}
    \caption{Kinematics map of [OIII]$\lambda$5007 line emission. Left panel: 
Velocity map. Middle panel: The map of W90 parameter, which is equivalent to 
velocity dispersion. Right panel: The map of asymmetry index. The red cross represents 
the core (optical {\it Gaia} position).} 
    \label{figure-5}
\end{figure*}

\begin{figure*}

 \includegraphics[scale=0.45]{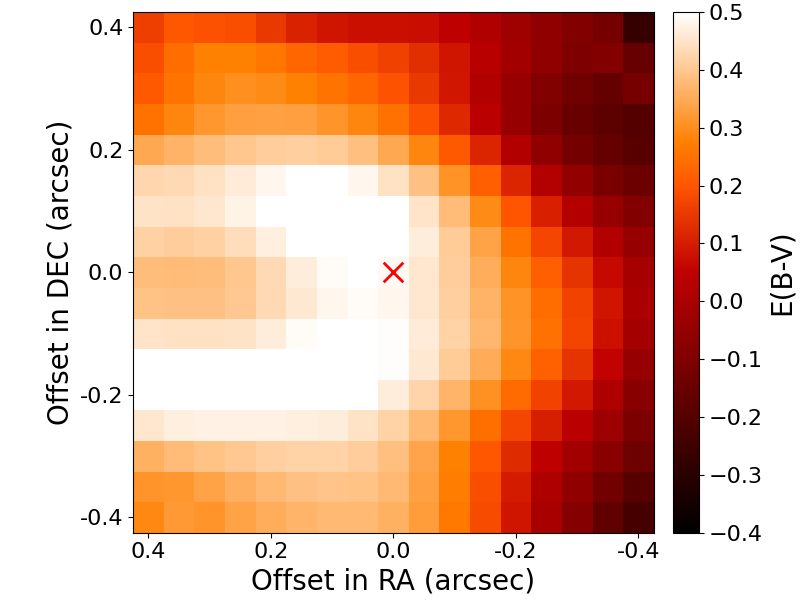}
 \includegraphics[scale=0.45 ]{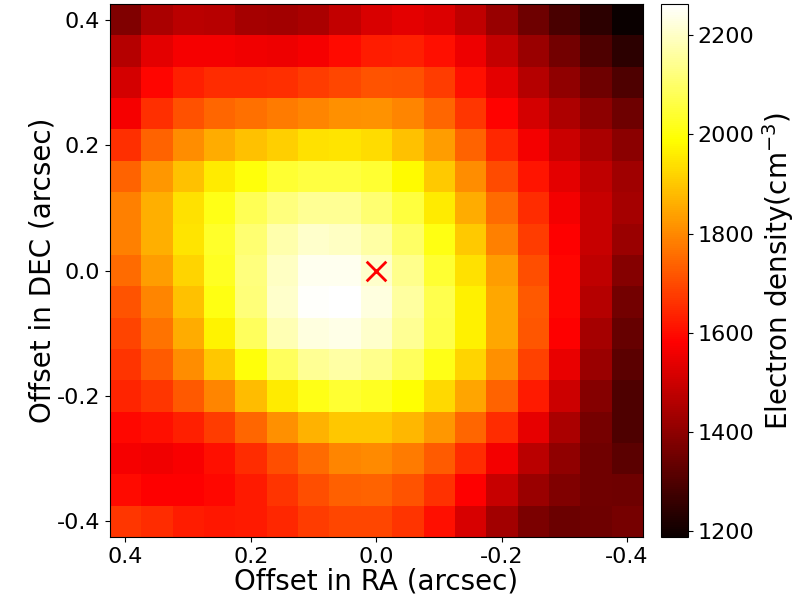}
     
    \caption{Left panel: The extinction E(B$-$V) map. Right panel: The map of 
electron density calculated from [SII]$\lambda\lambda$6717,6731 line ratio. The 
red cross is the core (optical {\it Gaia} position).} 
    \label{figure-6}
\end{figure*}

\begin{figure}
\includegraphics[scale=0.45]{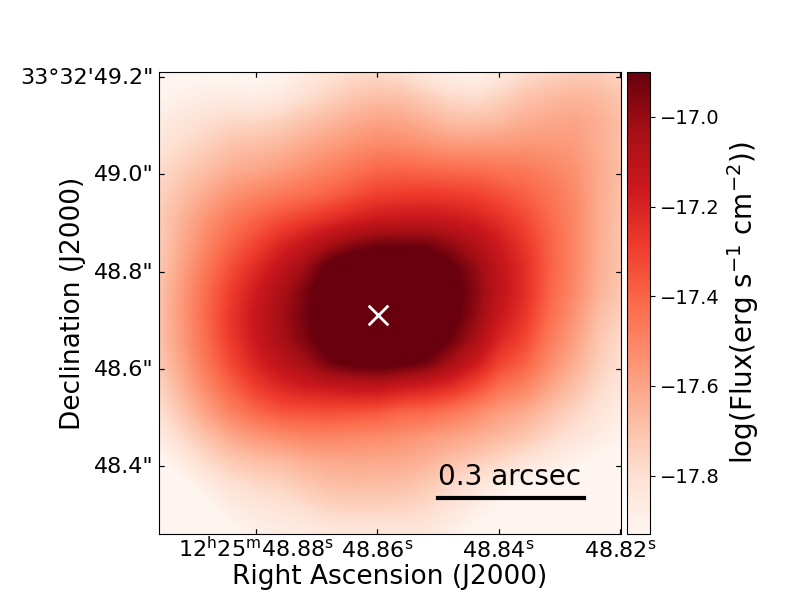}
\caption{The molecular H$_2$$\lambda$2.4085 image of NGC~4395. The core of NGC~4395 taken as the optical {\it Gaia} position is shown as a white cross.}
\label{figure-7}
\end{figure}

\begin{figure*}
    \centering
     \vbox{
    \hspace*{-1cm} \includegraphics[scale=0.4]{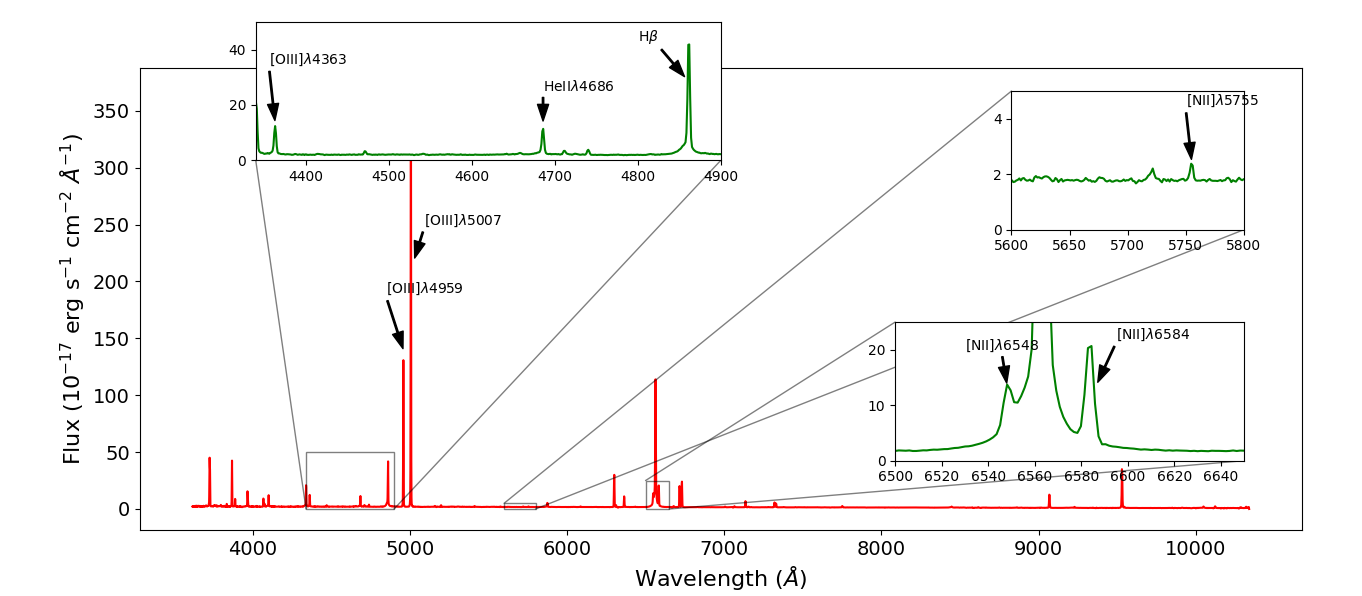}
    }
    \caption{ The rest frame spectrum of NGC~4395 from MaNGA covering the central brightest 0.5$^{\prime\prime}$$\times$0.5$^{\prime\prime}$.
    The lines used to calculate the line intensity ratios are marked.}
    \label{figure-8}
\end{figure*}

\begin{figure*}
\centering
\vbox{
     \hbox{
          \includegraphics[scale=0.45]{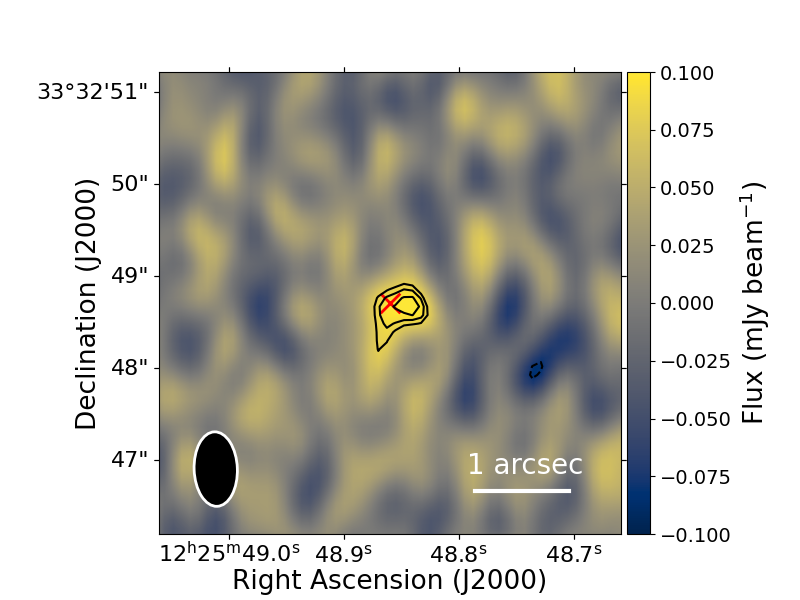}
          \hspace{-0.2 cm}
          \includegraphics[scale=0.45]{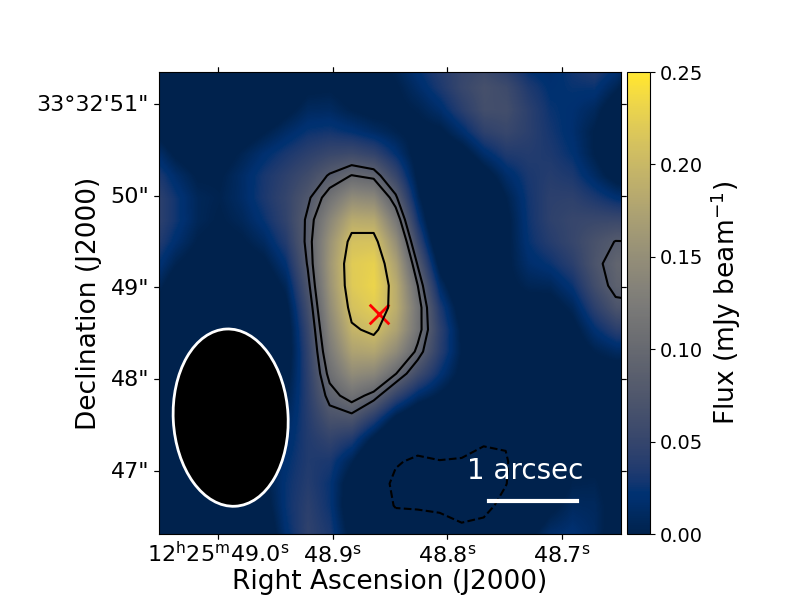}
          }
     \hbox{
           \includegraphics[scale=0.45]{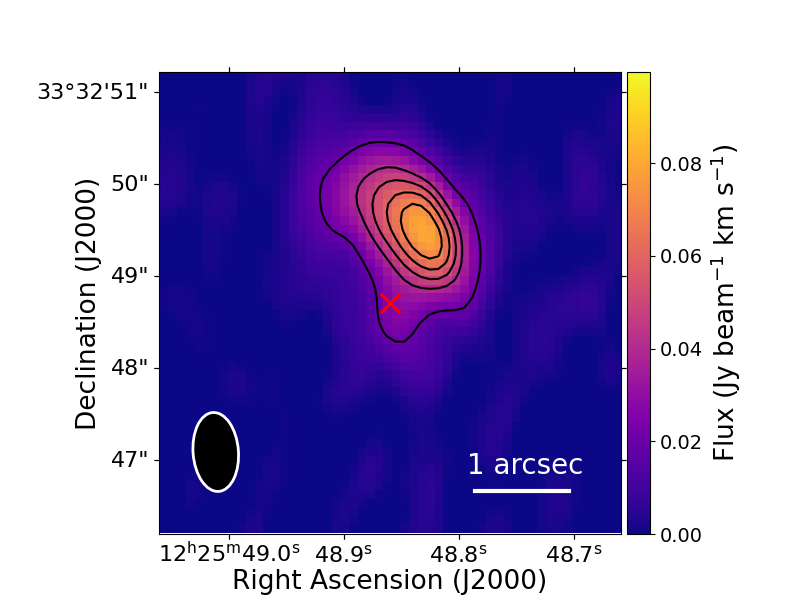}
           \includegraphics[scale=0.45]{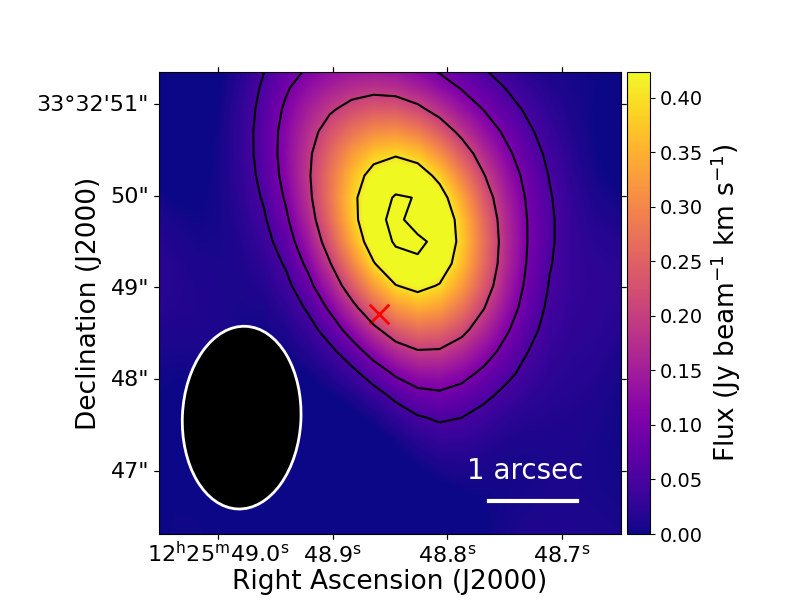}
          }
     }
\caption{Upper panel: The continuum emission at 237 GHz based on observations 
with ALMA on 22 March 2018 (left) and 23 January 2019 (right). The contours are 
-0.08, 0.08, 0.09 and 0.1 Jy/beam (left) and -0.08, 0.08, 0.1, and 0.2 Jy/beam (right). Lower 
panel: The CO(2$-$1) emission for the observation on 22 March 2018 (left) and 
23 January 2019 (right). The contours are 0.02, 0.04, 0.05, 0.06 and 
0.07 Jy beam$^{-1}$ km s$^{-1}$ (left) and 0.05, 0.1, 0.2, 0.4
and 0.5 Jy beam$^{-1}$ km s$^{-1}$ (right). The red cross is the optical {\it Gaia} position. The synthesized beams are shown by black ellipses.}
\label{figure-9}
\end{figure*}

\begin{figure*}
\hbox{
     \includegraphics[scale=0.45]{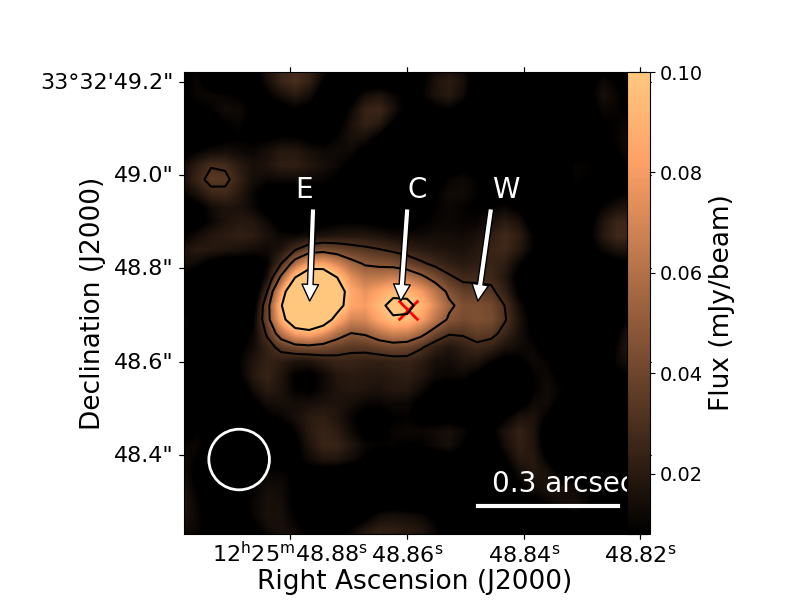}
     \includegraphics[scale=0.45]{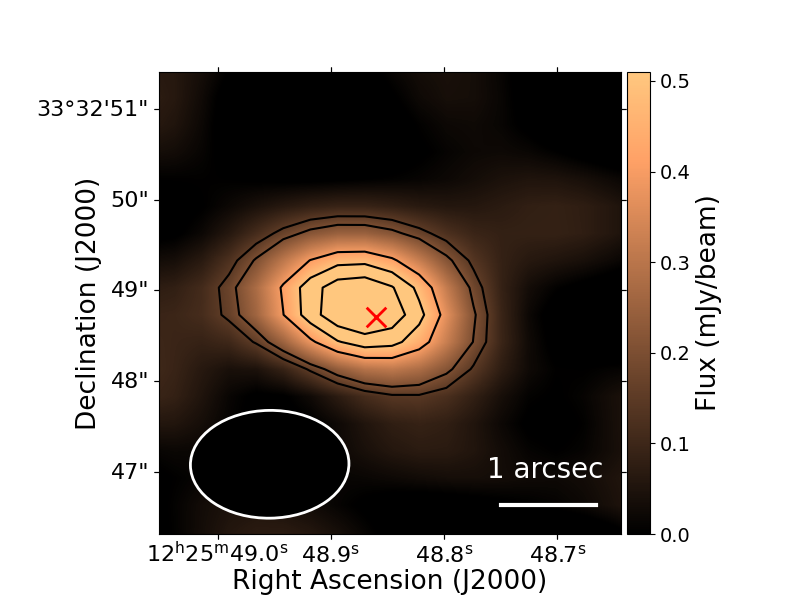}
     }
\caption{Left: The VLA radio images at 15 GHz with a size of 1$^{\prime\prime}$$\times$1$^{\prime\prime}$. The contours are at 0.03, 0.05, 0.10 mJy/beam. The rms noise is 11 $\mu$Jy/beam.
Here, C is the radio core, E is the eastern jet component and W is the 
western jet component. Right: The 4.8 GHz VLA image with a size of 5$^{\prime\prime}$$\times$5$^{\prime\prime}$. The contours are at 0.15, 0.20, 0.40, 0.50, 0.60 mJy/beam. The rms noise is 
48 $\mu$Jy/beam. The synthesised beams are shown as white ellipses. These have a size of
0.129$^{\prime\prime}$$\times$0.124$^{\prime\prime}$ with PA of $-$18 deg (left panel) and 1.75$^{\prime\prime}$$\times$1.19$^{\prime\prime}$ with PA of 89 deg (right panel). The red cross in both the panels represents the optical {\it Gaia} position.}
\label{figure-10}
\end{figure*}

\section{Data reduction and analysis}
To characterise the jet$-$ISM interaction on the scale of parsecs, we utilised data from
both ground and space-based telescopes from low energy radio waves to high energy 
X-rays. The details of the data sets used are given in Table \ref{table-1}.

\begin{table*}
    \centering
    \caption{Details of the data used in this work. These are all archival observations.}
%    \tiny
%    \begin{tabular}{p{0.09cm} p{0.25cm} p{0.1cm} p{0.25cm} p{0.12cm}}
    \begin{tabular}{lcccc}
    \hline
     Telescope  & Filter$/$Wavelength$/$Energy & FoV  &  Resolution$/$Synthesised Beam size &  Exposure time \\
     \hline
     Chandra    & 0.5$-$7.0 keV  & $\sim$ 8$^{\prime}$  & $\sim$ 1$^{\prime\prime}$ & 79ks \\
     HST        & F507N ($\lambda_{eff}$=5009.7~\AA, $\Delta\lambda$=66.8~\AA) & $\sim$ 2.7$^\prime$ & $\sim$ 0.1$^{\prime\prime}$ &  1521s \\
                & F547M ($\lambda_{eff}$=5435.8~\AA, $\Delta\lambda$=717.0~\AA)  & $\sim$ 2.7$^{\prime}$ & $\sim$ 0.1$^{\prime\prime}$ & 746s \\
     MaNGA      & 3600~\AA$-$10000~\AA   & $\sim$ 30$^{\prime\prime}\times$ 30$^{\prime\prime}$ & $\sim$ 1.5$^{\prime\prime}$ & 5400s\\       
     GMOS       & 4500~\AA$-$7300~\AA  & $\sim$ 5.0$^{\prime\prime}\times$3.5$^{\prime\prime}$  &  $\sim$ 0.5$^{\prime\prime}$ & 2880s\\
     NIFS       & 2.2 $\mu$m  & $\sim$ 3.4$^{\prime\prime}\times$3.4$^{\prime\prime}$  &  $\sim$ 0.2$^{\prime\prime}$ & 5400s \\
     VLA        & 5 GHz       & $\sim$ 8$^{\prime}$  & 1.75$^{\prime\prime} \times$1.19$^{\prime\prime}$, 89 deg  & 1090s  \\
                &  15 GHz     & $\sim$ 3$^{\prime}$ & 0.129$^{\prime\prime}\times$0.124$^{\prime\prime}$, $-$18 deg & 239s \\
     ALMA       & 237 GHz, CO(2$-$1)  & $\sim$ 19$^{\prime\prime}$ & 1.94$^{\prime\prime}\times$1.25$^{\prime\prime}$, 3 deg  & 2037s   \\
                & 237 GHz, CO(2$-$1)  & $\sim$ 19$^{\prime\prime}$ & 0.81$^{\prime\prime}\times$0.47$^{\prime\prime}$, 356 deg  &423s   \\
     \hline
    \end{tabular}
    \label{table-1}
\end{table*}

\subsection{X-ray}
We used four epochs of observations (OBSID: 402, 882, 5301, 5302) carried out by 
the {\it Chandra} X-ray observatory with the advanced CCD imaging spectrometer (ACIS,
0.5$-$7 keV) for exposures ranging from $\sim$ 1 ks to $\sim$ 31 ks.
We reduced the data using the Chandra Interactive Analysis
of Observations (CIAO, version 4.14) software and calibration files
(CALDB version 4.9.8). We first downloaded the data and reprocessed them
by running the task {\tt chandra\_repro} to generate the cleaned and
calibrated event files. Next, we  combined all the event files, 
computed the exposure maps and generated 
an exposure-corrected image in the default 0.5$-$7 keV energy range for a
total exposure of $\sim$ 79 ks. We adopted the task {\tt merge\_obs} 
for this purpose. We also rebinned the data by one-quarter of the native
0.492$^{\prime\prime}$ per pixel giving an effective resolution of
0.123$^{\prime\prime}$ per pixel. The image with a size of 1$^{\prime\prime}$$\times$1$^{\prime\prime}$ is shown in Fig. \ref{figure-1}. The 
red cross in the image is the optical {\it Gaia} position.

\subsection{Optical Imaging}
Observations of NGC~4395 carried out by HST WFC3-UVIS2 using a range of filters are available in the HST archives\footnote{https://archive.stsci.edu/} {(Proposal ID: 12212, PI: D. Michael Crenshaw)}. Of these, we used the data 
in two filters, one F502N, which is centred at 5009.87 \AA ~and the other F547M, which is centred around the 
nearby continuum at 5756.9 \AA.
%We converted the observed [OIII]$\lambda$5007 F502N and F547M images to flux  scale using the KEYWORD {\tt PHOTFLAM} given in the image headers. 
Our aim here is to get the [OIII]$\lambda$5007 image from the observation done in 
the F502N filter, which in addition to the [OIII]$\lambda$5007 line emission 
also contains the continuum emission. One of the ideal ways to remove the continuum 
emission from the observed F502N narrow-band image is to use observations in 
medium band filters both blue-ward and red-ward of the F502N filter. 
In such a scenario, observations in the medium band filters that flank
the narrow band filter can be used to get the continuum slope. The slope thus obtained can give us an estimate of the continuum that can be subtracted from the F502N data to get the line image. However, 
in the present case, we have only one medium band filter observation which is adjacent and redward of the F502N filter. Given this, we adopted the following to get the line image. 
Under the assumption that the continuum emission around [OIII]$\lambda$5007 line has a zero slope (which is in agreement with spectra in this case), we selected six source-free regions having sizes of 10$^{\prime\prime} \times$ 10$^{\prime\prime}$ in the F547M and F502N observations, determined the scale factors and applied the mean scale factor (c) and subtracted one filter observation from the other as given below
\begin{equation}
    f(line) = c*f(N) - f(M)
\end{equation}
Here, f(N) and f(M) are the brightness in counts/sec in the narrow F502N filter 
(that contains the [OIII]$\lambda$5007 emission line) and medium band filter 
F547M (that contains the continuum emission). The parameter c is the scaling factor, 
which is the mean of the ratio of the fluxes  in the narrow-band filter, F502N 
and the medium-band filter, F547M determined from six source free regions. Then we converted the observed [OIII]$\lambda$5007 (continuum subtracted) images to flux  scale using the KEYWORD {\tt PHOTFLAM} given in the image headers. The observed medium-band F547M image, the narrow-band F502N image and the continuum-subtracted image are shown in Fig. \ref{figure-3}. These images have an FWHM angular resolution of $\sim$ 0.1$^{\prime\prime}$.

\subsection{Optical/infrared integral field spectroscopy}
We used archival near infrared and optical integral field spectroscopic (IFS) 
observations obtained with the Gemini and SDSS telescopes. 

\subsubsection{Gemini}
For the optical, we used the archival data from the Gemini Multi-Object
Spectrograph (GMOS) under the program ID GN-2015A-DD-6 (PI. Mason Rachel). GMOS 
with a field of view (FoV) of 5.0$^{\prime\prime}$$\times$3.5$^{\prime\prime}$ covers 
the spectral 
range from 4500$-$7300 \AA.  In the infrared, we used the archival data from the 
adaptive optics assisted K-band observations acquired with the near infrared 
integral field spectrograph (NIFS) under the program ID GN-2010A-Q-38 (PI. Anil 
Seth). The K-band centered at 2.2 $\mu$m covers an FoV of 3.4$^{\prime\prime}$$\times$3.4$^{\prime\prime}$. The data cubes in GMOS observations have a spatial sampling of 
0.05$^{\prime\prime}$/pixel. The spectral resolution is 
90 km s$^{-1}$ and the angular resolution is 0.5$^{\prime\prime}$ \citep{2019MNRAS.486..691B}. Similarly, 
the data cubes in the NIFS observations have a spatial sampling of 
0.05$^{\prime\prime}$/pixel. The spectral resolution is 45 km s$^{-1}$ 
and the angular resolution is 0.2$^{\prime\prime}$ \citep{2019MNRAS.486..691B}.
We reduced the GMOS and NIFS data following standard procedures in IRAF (see 
\citealt{2019MNRAS.486..691B} for details). The GMOS data is seeing limited, 
while NIFS data is AO-assisted.

For fitting the emission lines, we followed a non-parametric approach 
which involves the removal of the underlying continuum and fitting multiple Gaussian components 
to the emission lines. In 
the case of GMOS data, we identified line-free regions on either side of our 
region of interest, namely the [OIII]$\lambda$5007 region 
($\lambda\lambda$ = 4990$-$5040 \AA).  We fitted a first-order polynomial to the line-free 
 regions and then subtracted the function from the observations. After 
continuum subtraction, we fitted the [OIII]$\lambda$5007 emission line with two 
Gaussian components to extract the flux and other properties of the line using
the non-linear least square minimization algorithm within {\tt Curvefit} module
of {\tt Scipy} \citep{2020SciPy-NMeth}.
An example of the fit is shown in Fig. \ref{figure-4}. For the broad 
 H$\beta$ line, we fitted three Gaussian components, one for the narrow AGN 
component, the second for the  broad AGN component and the third for the  
broad outflowing component. While fitting the broad and narrow AGN components 
we fixed the peak velocity to be the same, however, the line widths were
treated as  free parameters and were allowed to vary. For the broad outflowing 
component, no restriction was imposed, either for the peak of the component or
the width of the component. From our fits to the H$\beta$ line, we
obtained a $\sigma$ of 385 kms$^{-1}$ for the broad H$\beta$ component, which is 
similar to the value of $\sigma$ = 334 kms$^{-1}$ obtained by 
\cite{2019MNRAS.486..691B}. The procedure adopted in this work to 
fit the emission lines is thus appropriate. We fitted the lines [NII]$\lambda\lambda$6548,6584 
and H$\alpha$ lines together. For the H$\alpha$ line we used three Gaussian
components similar to that used for the H$\beta$ line. In addition, we used 
two Gaussians for the two [NII]$\lambda\lambda$6548,6584 lines. Here, we tied 
the widths of the [NII]$\lambda\lambda$6548,6584 lines to the width of the narrow 
component of H$\alpha$. For the narrow lines such as [NII]$\lambda$5755, 
[SII]$\lambda\lambda$6716,6732 and the H2$\lambda$2.4085 line from NIFS, we 
followed the same methodology used for the  [OIII]$\lambda$5007 line.

From the Gaussian fits to the [OIII]$\lambda$5007 line emission in the 
observed spectra (not corrected for instrumental resolution), we estimated 
non-parametric values \citep{2014MNRAS.442..784Z} such as the  velocity ($v50$, 
the velocity where the cumulative flux of the line becomes half of the total flux), 
velocity dispersion $(W90 = v95 - v5$, where $v95$ and $v5$ are the velocities 
at which the flux becomes 95$\%$ and 5$\%$ of total flux) and the asymmetry 
$\left(R= \frac{(v95 - v50)-(v50 - v5)}{v95-v5}\right)$ of the line. Also, from 
fits to the [SII] doublet and using the ratio of the [SII]$\lambda$6716 to 
[SII]$\lambda$6731 lines, we estimated the electron density. This line ratio is 
sensitive to electron densities of the order of $\sim$ 10$^2$ - 10$^4$ cm$^{-3}$. 
For estimating the electron density, we assumed an electron temperature 
of 10,000 K.
We calculated the internal extinction E(B$-$V) from H$\alpha$ and H$\beta$ line ratio 
using the following formula \citep{1972ApJ...172..593M, 1995ApJS...98..171V}
\begin{equation}
\label{eq:extinction}
E(B-V)= 1.925 \times log{\frac{\left(\frac{I_{H\alpha}}{I_{H\beta}}\right)_{obs}}{3.1}}.
\end{equation}
The maps for the velocity and velocity dispersion of the [OIII]$\lambda$5007 line 
emitting gas and for the asymmetry parameter of the line are given in 
Fig.~\ref{figure-5}, whereas the E(B-V) map and the electron density map are 
shown in Fig.~\ref{figure-6}. The molecular H$_2$$\lambda$2.4085 image is shown in Fig. 
\ref{figure-7}. These figures cover approximately central 1$^{\prime\prime} 
\times$1$^{\prime\prime}$ region of NGC~4395. This is because 
the radio emission at 15 GHz has an extension of $\sim$ 0.5$^{\prime\prime}$ and
our aim is to investigate the interaction of the radio jet with the ISM.

\vspace{3 pt}
\subsubsection{ SDSS/MaNGA}
From the Sloan Digital Sky Survey (SDSS) we  used data of NGC~4395 observed
as part of the Mapping Nearby Galaxies at Apache Point Observatory (MaNGA) 
survey.  The pixel scale of MaNGA product is 0.5$^{\prime\prime} \times$0.5$^{\prime\prime}$. It covers the wavelength range of 3600 \AA ~to 10000 \AA ~with a spectral 
resolution ($\lambda/\Delta\lambda$) of $\approx$ 2000. We used the spectrum of 
the central brightest pixel which covers the central 0.5$^{\prime\prime} 
\times$0.5$^{\prime\prime}$ region of NGC~4395, and largely encompasses the 
$\sim$ 0.5$^{\prime\prime}$ extent (total) of the radio jet, the region of interest in this paper 
to investigate the radio jet$-$ISM interaction. By using the 
advantage of this wavelength region, we detected shock sensitive lines 
[OIII]$\lambda$4363, HeII$\lambda$4886 (which are beyond the limit of GMOS), 
[OIII]$\lambda \lambda$4959,5007,  H$\beta$, [NII]$\lambda$5755 and 
[NII]$\lambda \lambda$6548,6584 lines (as shown in Fig \ref{figure-8}) and 
estimated the parameters. These parameters are thus a representation of 
the physical conditions in the central 0.5$^{\prime\prime} \times$0.5$^{\prime\prime}$ 
region.  We fitted the emission lines in the 
same way as explained in Section 2.3.1  and estimated the emission line fluxes.

\subsection{ALMA}
We used the archival data, observed with the high-resolution 
Atacama Large Millimeter/submillimeter Array (ALMA) with 12-m antennas (Data-ID: 
2017.1.00572.S, PI: Davis, Timothy). 
The observations were carried out on March 22, 2018, and January 23, 2019, with ALMA band 6 in the frequency 
range of 227.47$-$246.43 GHz. The on-source integration time was 2037 s and 423 s, respectively. During 
the observations, a total of 46 antennas were used, with a minimum baseline of 
15.1 m and a maximum baseline of 783.5 m. For the observations on both days, the quasar J1221+2813 was observed as 
a phase calibrator, and J1229+0203 was observed as a flux density and bandpass calibrator.

We reduced the data using the Common Astronomy Software Application (CASA v5.4.1) 
with the standard data reduction pipeline of the ALMA observatory. 
We show in Fig.~\ref{figure-9} (upper left) the continuum image of NGC~4395 at 
237.1227 GHz observed on 22 March 2018 having a synthesized beam size of 
0.805$^{\prime\prime}$$\times$0.469$^{\prime\prime}$ along position angle (PA) of 356 deg.  From two-dimensional Gaussian fits we found the peak and integrated flux densities to 
be 93$\pm$20 $\mu$Jy beam$^{-1}$ and 131$\pm$45 $\mu$Jy respectively. 
Also, the continuum image at 237.1227 GHz, observed on 23 January 2019 is shown 
in Fig.~\ref{figure-9} (upper right). It has a synthesized beam size of 
1.935$^{\prime\prime}$$\times$1.253$^{\prime\prime}$ along PA of 3 deg. From Gaussian fits to the data, we found the peak and integrated flux densities to be 274$\pm$21 $\mu$Jy beam$^{-1}$ and 287$\pm$41 $\mu$Jy respectively. These results from an independent analysis are also in agreement with those of \cite{2022MNRAS.514.6215Y}. We used the task {\tt TCLEAN} to generate the spectral data cubes. The CO line maps are shown in Fig. \ref{figure-9} (lower panels). From both the observations, we found the peak of the CO(2$-$1) emission to be displaced by around 0.9$^{\prime\prime}$ ($\sim$ 20 parsec) from the nucleus (as determined by \textit{Gaia}) of NGC~4395.

\subsection{VLA}
The source was observed with the VLA A-configuration at 15GHz (PI: Payaswini Saikia, 
Legacy  ID: AS1409). We reduced the data using  standard  procedures that include
flagging bad data  using  
{\it CASA} (see \citealt{2018A&A...616A.152S} for  details). The  beam  size  obtained  
is  0.129$^{\prime\prime}$$\times$0.124$^{\prime\prime}$  with a PA of $-$18 deg. The 
final image at 15 GHz has an rms noise of 11 $\mu$Jy/beam. The contours 
of the 15-GHz image shown in the left panel of Fig. \ref{figure-10} 
are at 0.03, 0.05 and 0.10 mJy/beam. The western jet component, has a peak flux density of 44.6$\pm$1.0 $\mu$Jy/beam. Though it is fainter related to the core and eastern components, it is detected at about the 4$\sigma$ level.

The source was also  observed  at  
4.8  GHz (C band) in the VLA  B-configuration  (PI:J.S.  Ulvestad,  Legacy  ID: 
AU079). We reduced this data using standard procedures in {\it AIPS} by using 3C286 
as the flux density calibrator and 1227+365 as a phase calibrator. We achieved an rms noise of 
48 $\mu$Jy/beam. The beam size in the final reduced image is 1.75$^{\prime\prime}$$\times$1.19$^{\prime\prime}$ along a position angle of  89 deg. The final images at 15 GHz and
4.8 GHz are shown in Fig. \ref{figure-10}.

\section{Result and Discussion}
\subsection{Radio morphology}
The VLA 
15~GHz image (Fig. \ref{figure-10}, left panel) showed the source to be a 
triple with E being the eastern component 
of the triple \citep{2018A&A...616A.152S}. The weaker central component of the 
triple is coincident with the optical \textit{Gaia} position, while component E is 
displaced from it by about 220 mas, corresponding to a projected distance 
of 4.6 parsec. The western component (W) is separated from the central component by 
4.2 parsec. However, the overall projected extension of the source is $\sim$ 10 parsec. The 
source is also highly asymmetric in brightness, the ratio of peak brightness of 
components E to W is 3.8. Component E has a spectral index, $\alpha$, 
of $-0.64\pm0.05$ (S$\propto \nu^{\alpha}$) and a brightness temperature, 
$T_B$ of (2.3$\pm$0.4)$\times$10$^6$ K, showing it to be a non-thermal source. 
For the central feature $\alpha = -0.12\pm0.08$, and non-detection of a 
sub-parsec scale compact component sets $T_B$ $<$ 5.9$\times$10$^5$ K 
\citep{2022MNRAS.514.6215Y}. Radio cores being resolved out in low-mass AGN 
when observed with milliarcsec resolution has been reported 
earlier \citep{2017ApJ...845...50N}. Variability or episodic nuclear jet activity 
could also contribute to the non-detection of a core. 

The triple structure is 
strongly reminiscent of bipolar jet ejection in radio-loud AGN, and we suggest that the outer components (W and E) are formed by weak radio jets from the intermediate-mass black hole, and refer to the central component as the radio core. We refer to W and E, the end-points of the radio emission as jets in this paper to explore jet$-$ISM interaction. Low power radio jets (P $<$ 10$^{42}$ erg s$^{-1}$) can have a significant effect on the ISM of the host galaxy, interacting with clouds of gas and heating the gas, entraining ambient gas, losing collimation and sometimes forming arc-like fronts \citep{10.1093/mnras/sty390}. 

\begin{figure}
    \centering
    \includegraphics[scale=0.4]{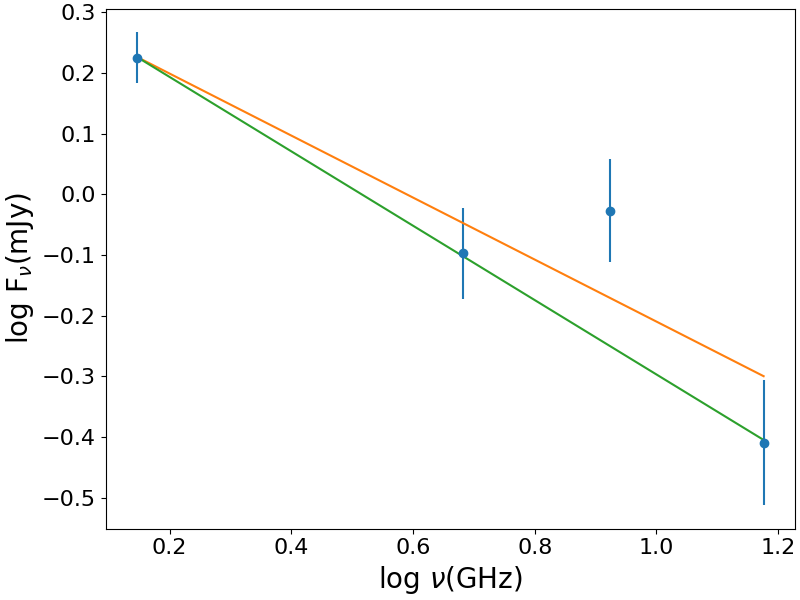}
    \caption{Plot of flux density against frequency. The solid lines are linear least squares fits to the data. The orange line is the fit to all the data points, while the green line is the fit excluding the 8.4 GHz measurement.}
    \label{fig:spectralindex}
\end{figure}

\subsection{Nature of radio emission in the central 10 parsec region}
The radio emission observed in the central parsec-scale region of a dwarf 
galaxy could be from a variety of physical processes, such as low-power jets, 
AGN-driven wind, SF, coronal activity and free-free emission from thermal 
gas \citep{2019NatAs...3..387P}. Radio structure, spectral index, polarization 
characteristics of the radio emission if detectable in future, spatial 
correlation of radio structure and different gaseous components, spectral 
line diagnostics of the different components of the ISM could provide 
valuable clues in identifying the dominant processes.
In NGC~4395, the radio morphology of a triple radio source clearly shows signs of interaction on the eastern side which appears to bend sharply, as seen in the high-resolution image \citep[see][]{2022MNRAS.514.6215Y},
indicating jet$-$ISM interaction. This was reinforced by a detailed study of the line-emitting gas.
The [OIII]$\lambda$5007 emission appears closely associated with the radio source, ionized by shocks as the jets flow outwards. Line-ratio diagnostics, estimation of gas temperatures from line ratios, and comparison of line ratios with theoretical predictions using {\tt MAPPINGS} models, all showed shocks to be the dominant process responsible for the ionization. The eastern component which is more prominent at radio wavelengths showed stronger signs of interaction with the ISM than the western component.

The spectral index (inclusive of all the three 
components) derived over the 1.4$-$15 GHz range (see Fig. \ref{fig:spectralindex}) gives a value of $\alpha$ = $-$0.51$\pm$0.11 (S$_\nu$ $\propto$ $\nu^{\alpha}$) and $\alpha$ = $-$0.61$\pm$0.01 excluding 8.4 GHz point. We note here that the flux density measurements are not simultaneous and could be affected by the variability of the central AGN.
However, considering only the eastern jet/component, \cite{2022MNRAS.514.6215Y} obtained a value of  $\alpha$ = $-$0.64$\pm$0.05. This is very close to the theoretical injection spectral index \citep{2000ApJ...542..235K}.
Considering the source to have a spectral index in the range $-$0.51 to $-$0.64, the inverse Compton scattering of the CMB photons and radio photons can give rise to a power law X-ray spectrum whose photon index, $\Gamma$ can be 1.51 to 1.64. This is close to the value of $\Gamma$ = 1.67 found by \cite{2019ApJ...886..145K} from an analysis of XMM data in the 2$-$10 keV band.

Using the observed luminosity at 1.4 GHz, we calculated 
the  jet power as \citep{2010ApJ...720.1066C}
\begin{equation}
P_{jet} = 5.8 \times 10^{43} \left(\frac{L_{1.4GHz}}{10^{40} erg s^{-1}} \right)^{0.7}
\end{equation}
Using L$_{1.4 GHz}$ = (6.1$\pm$0.3)$\times$10$^{34}$ erg s$^{-1}$ \citep{2001ApJS..133...77H} we 
 estimated the jet power to be 
P$_{jet}$ = (1.3$\pm$0.3)$\times$10$^{40}$ erg s$^{-1}$.
It is thus evident that the jet in NGC~4395 is weak compared to powerful radio 
galaxies having typical jet powers of 10$^{44}-10^{45}$ erg s$^{-1}$ \citep{10.1093/mnras/stw3330}. 

The above considerations all show that the radio emission in the central 10 parsec region in NGC~4395 is from a low-power jet launched by an intermediate-mass black hole.

\subsection{Radio and [OIII]$\lambda$5007 emission}
\begin{figure*}
\hbox{
    \centering
     \includegraphics[scale=0.26]{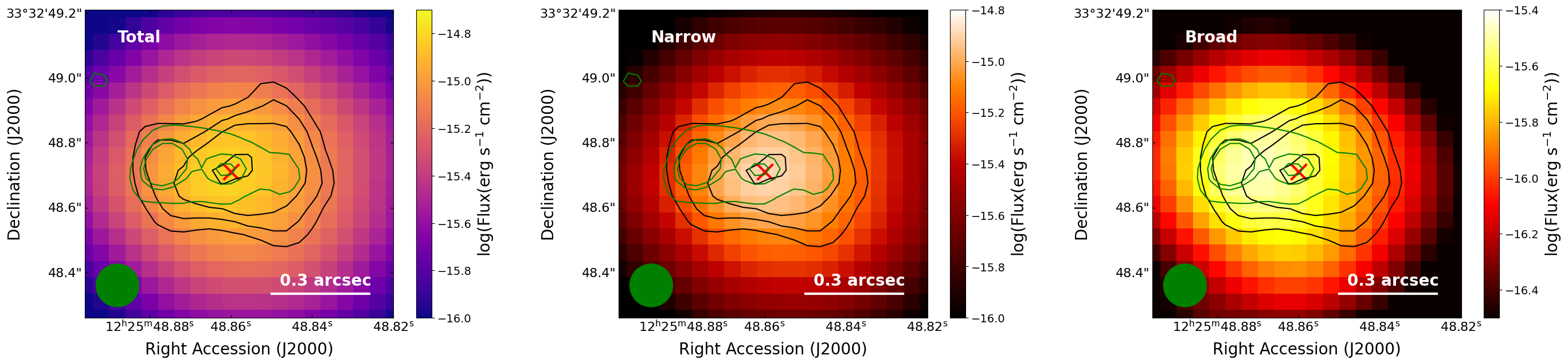}
     }
    \caption{GMOS [OIII]$\lambda$5007 image in colour scale along 
with the radio emission at 15 GHz (green contours) with contour levels of 0.03, 
0.08, 0.1 mJy/beam  and HST [OIII]$\lambda$5007 emission (black contours) with contour 
levels of (0.05, 0.1, 0.2, 1.0)$\times$10$^{-15}$ erg s$^{-1}$ cm$^{-2}$. The left 
panel shows the GMOS total flux in [OIII]$\lambda$5007, the middle panel shows 
the GMOS flux in the narrow component of [OIII]$\lambda$5007 and the right hand 
panel shows the GMOS flux in the broad component of [OIII]$\lambda$5007. The red 
cross is the core (optical {\it Gaia} position). The green circle is the 
synthesised beam of 15 GHz data with a size of 0.129$^{\prime\prime}$$\times$0.124$^{\prime\prime}$ along PA $-$18 deg.}
    \label{figure-11}
\end{figure*}

Fig.~\ref{figure-11} shows the [OIII]$\lambda$5007 map of NGC~4395 over a region 
of 1$^{\prime\prime}$$\times$1$^{\prime\prime}$ in the total line emission (left panel), the narrow line component (middle panel) and the broad outflowing line component (right panel). Also, overplotted in these figures are the 15 GHz radio contours in 
green and the [OIII]$\lambda$5007 HST emission in black. The broad outflowing component of [OIII]$\lambda$5007 emission is brighter in the eastern side, where 
the radio emission also tends to be brighter. From these figures, it is evident that the total [OIII]$\lambda$5007 emission is prevalent over the entire extent 
of the radio emission, with the peak of the total [OIII]$\lambda$5007 emission 
coinciding with the peak of the central 15 GHz emission and the outflowing component of the [OIII] line emission is displaced towards the east and overlaps with the eastern radio component. We note here that the 
peak of the [OIII]$\lambda$5007 flux from GMOS and that from HST are comparable, with values of (1.56$\pm$0.08)$\times 10^{-15}$ erg s$^{-1}$ cm$^{-2}$ and
(1.25$\pm$0.06)$\times 10^{-15}$ erg s$^{-1}$ cm$^{-2}$ respectively. The pixel scales of the [OIII]$\lambda$5007 images 
from GMOS and HST are 0.05$^{\prime\prime}$ and 0.04$^{\prime\prime}$, respectively. The HST 
[OIII]$\lambda$5007 emission is asymmetric, has a cone-like structure with a 
convex shape near the terminal points of the radio jet, resembling bow shocks in FRII radio galaxies \citep{1999ASPC..176..377K}. 
The eastern [OIII]$\lambda$5007 cone has a narrower opening angle, while the 
western cone has a wider opening angle. 
Also, the western [OIII]$\lambda$5007 cone has a more pronounced convex shaped morphology compared with the eastern one.
The asymmetry in the [OIII]$\lambda$5007 emission could be due to the 
jets passing through an inhomogeneous medium. The eastern jet is 
brighter in the radio band, travelling in a denser and larger E(B$-$V) 
medium (see Fig. \ref{figure-6}), could have ionized the gas, leading 
to dimmer [OIII]$\lambda$5007 emission and brighter synchrotron emission because 
of shock-induced compression.  Similarly, the western jet is travelling in a 
less dense medium, with smaller E(B$-$V) values, and enhanced [OIII]$\lambda$5007 
emission. The observed morphology of the source in radio and [OIII]$\lambda$5007 
is unambiguous evidence for the interaction of the radio jets with the ISM of 
the host of NGC~4395 and is the first structural evidence of jet$-$ISM interaction 
operating on scales $\sim$ 10 parsec in a triple radio source. 

\subsection{Multi-wavelength structure of NGC~4395}
The 15 GHz image (Fig.  \ref{figure-10}) shows the highly 
asymmetric triple structure discussed earlier. In luminous radio galaxies, the 
components seen on the side of the jets that interact with a dense cloud in the 
ISM are usually nearer and brighter \citep{2021A&ARv..29....3O,2003PASA...20...50S}, 
as there is greater dissipation of energy on this side and the dense clouds 
inhibit the advancement of the jets. In the case of NGC~4395, the brighter 
component is farther from the nucleus, although its high-resolution radio 
structure and our optical emission line study suggest the interaction of the jet 
with the ISM. Therefore a degree of intrinsic asymmetry in the radio jets 
cannot be ruled out. The velocities of the jets in these low-luminosity AGN are small so that relativistic beaming effects are not expected to be important, as also seen in the case of Seyfert galaxies \citep[e.g.][]{2000evn..conf....7R,2004IAUS..222..299W,2011ApJ...731...68L}.

Fig. \ref{figure-3} shows the [OIII]$\lambda$5007 image 
of the 1$^{\prime\prime}$$\times$1$^{\prime\prime}$, from HST. We found the [OIII]$\lambda$5007 
emission to be prevalent over the central 1$^{\prime\prime}$$\times$1$^{\prime\prime}$. The 
central component of the 15 GHz radio emission and the peak of the [OIII]$\lambda$5007 image
from HST,  coincides with the optical {\it Gaia} position. The {\it Gaia} position 
is thus the AGN core. The right-hand panel of Fig. \ref{figure-3} clearly shows 
that the [OIII] gas distribution is asymmetric around the optical {\it Gaia} 
position (the AGN core). Also, the [OIII] emission exhibits a convex shaped 
morphology at the terminal points on either side of the AGN core, though less
conspicuous on the eastern side, but more prominent on the western side 
where the radio emission is weaker. These features suggest that the structure 
seen in [OIII] line emission could be an ionised outflow, which we explore later. Similar conclusion was also arrived at by \cite{2019NatAs...3..755W}.

In Fig. \ref{figure-7} we show the 1$^{\prime\prime}$$\times$1$^{\prime\prime}$ map of the source in molecular H$_2$ at $2.4085$ $\mu$m, obtained from
NIFS on the Gemini telescope. The molecular H2$\lambda$2.4085 is also extended, 
in the East-West direction and spatially coincident with the 15 GHz radio emission.
The 4.8 GHz emission (Fig. \ref{figure-10}, right panel) is also spatially coincident 
with the 15 GHz emission and oriented in the East-West direction. The continuum 
emission at 237 GHz too (the top panels in Fig. \ref{figure-9})  coincides 
with the central radio source at 15 GHz and 
the optical {\it Gaia} position. However, the CO(2$-$1) line emission is 
concentrated at a larger distance ($\sim$ 0.9$^{\prime\prime}$) from the central nuclear 
emission (see Fig. \ref{figure-9}). The X-ray image (Fig. \ref{figure-1}) 
too shows emission centred around the nuclear 
region and having extended emission along the East-West direction.

\subsection{BPT analysis}
\begin{figure*}
\hbox{
    \centering
\hspace*{-0.3cm}\includegraphics[scale=0.38]{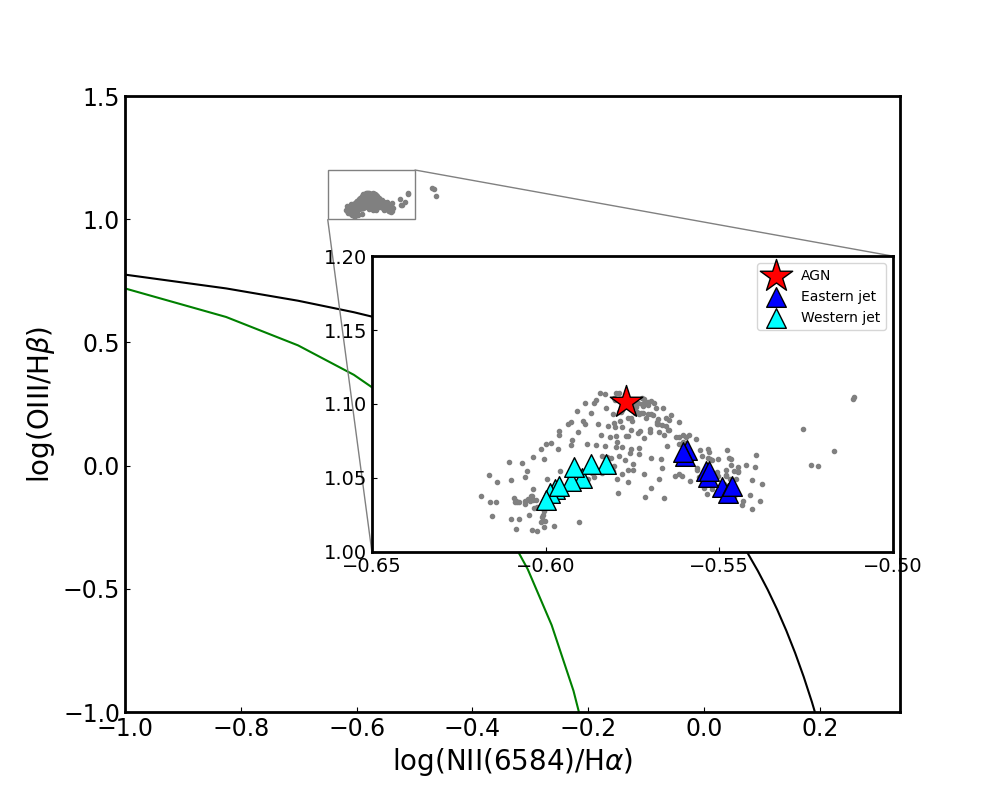}
\hspace*{-0.6cm}\includegraphics[scale=0.38]{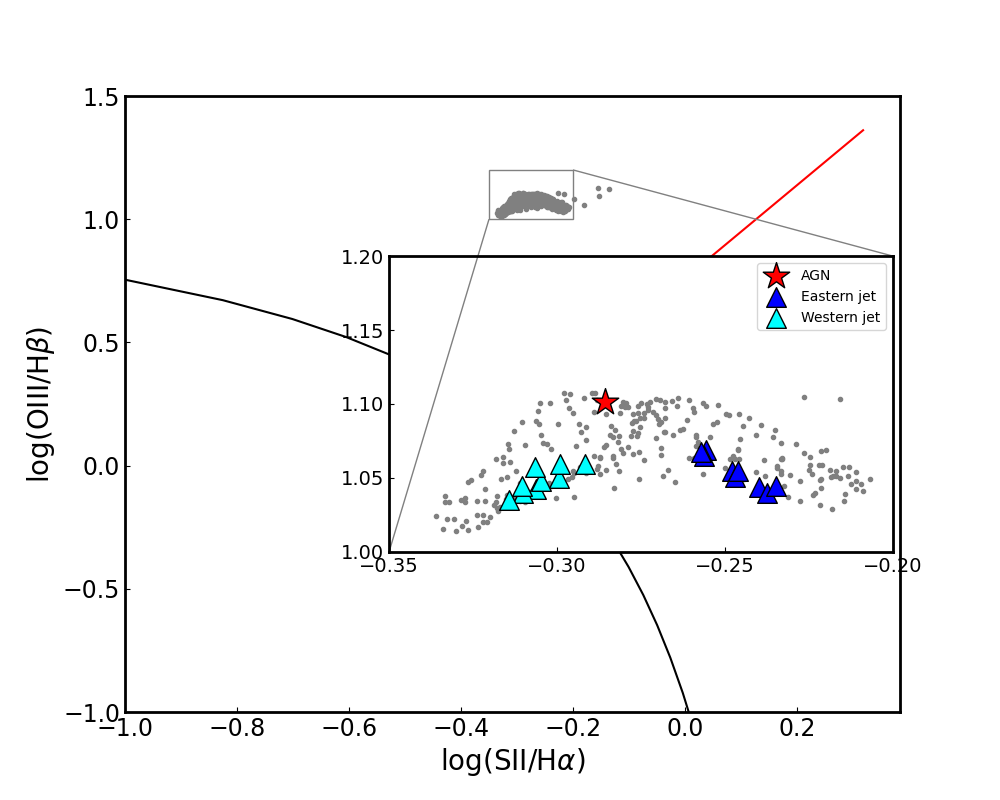}
     }
    \caption{The position of the spaxels belonging to the central 1$^{\prime\prime}$$\times$1$^{\prime\prime}$ region of NGC~4395 in the line ratio diagnostic diagrams. The green solid curve is from \cite{10.1111/j.1365-2966.2003.07154.x}, black and red solid lines are from \cite{2001ApJS..132...37K}. The typical error in these plots is 0.01 dex in both axes.}
    \label{figure-12}
\end{figure*}

Emission line ratios in the optical are an essential tool to distinguish between
star forming galaxies and AGN. Also, they can be used to disentangle
processes that lead to the line emission from SF, AGN and shocks. To measure the emission line fluxes, we fitted line profiles of
H$\alpha$ and [NII]$\lambda\lambda$ 6548,6584, [SII]$\lambda\lambda$ 6717,6731,
[OIII]$\lambda$5007 and H$\beta$ in the spectra of each spaxel, using two Gaussian
components for narrow lines and three Gaussian components for the broad Balmer lines H$\alpha$ and H$\beta$ 
(see Sect. 2.3.1). The extra component in all lines is to represent the contribution from outflowing gas, while other components are for the broad line region (BLR) and narrow line region (NLR). During the fitting of the [NII] and
H$\alpha$ lines, the line widths of the narrow components were tied together, and the peak fluxes were left free. For Balmer lines (H$\alpha$ and H$\beta$) we used the same velocity shift for the narrow component and one broad component, which are responsible for the NLR and BLR region, respectively. While fitting the [SII] lines, the width of these two lines was tied together.
We used the [OIII]$\lambda$5007/H$\beta$ versus [SII]/H$\alpha$ as well 
as [OIII]$\lambda$5007/H$\beta$ versus [NII]/H$\beta$ diagnostic diagrams to 
investigate the physical processes causing the emission lines. These diagnostic 
diagrams are shown in Fig.~\ref{figure-12}. Each point in these diagrams 
represents one spaxel in the 1$^{\prime\prime}$$\times$1$^{\prime\prime}$ region centered around 
NGC~4395. Here, the red star is the AGN, and the blue and cyan triangles 
represent the spaxels in the eastern and western jet components. Though  all the 
spaxels lie in the AGN region of the Baldwin, Phillips and 
Terlevich (BPT; \citealt{1981PASP...93....5B}) diagram, there is a clear 
segregation between the core, the eastern and the western components.

\subsection{Diagnostics of the emission lines: Photoionization by AGN and/or shock}
\subsubsection{Photoionization modelling }
To characterise the ionization processes that operate in the central 1$^{\prime\prime}$ region
of NGC~4395, we carried out a comparison of emission line measurements from the observed
GMOS spectra with photoionization and shock models from {\tt MAPPINGS-III} \citep{2013ascl.soft06008S} and implemented in {\tt ITERA} \citep{2013ascl.soft07012G}.
The emission lines in the spectra of material photoionized by AGN depend on the ionization parameter U, the slope of the ionizing continuum, $\beta$ ($\phi_\nu \propto \nu^{\beta}$),  the gas density and its metallicity.
We generated output spectra for a range of input parameters with $\beta$ ranging from $-$2 to $-$1.2 and log (U) varying from $-$4.0 to 0.0. We assumed solar metallicity \citep{2002A&A...391..809C} and a hydrogen density of $n_H$ = 1000 cm$^{-3}$.

\begin{figure*}
\hbox{
    \centering
\includegraphics[scale=0.45]{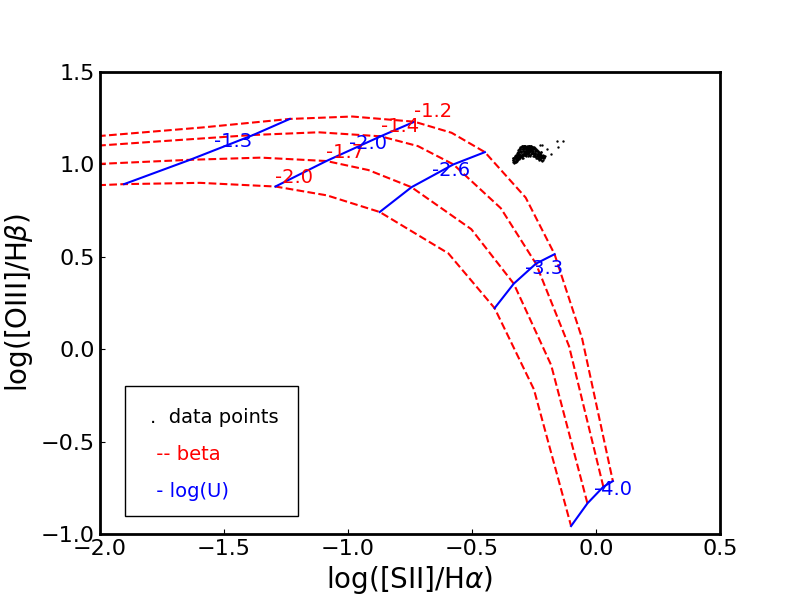}
\includegraphics[scale=0.45]{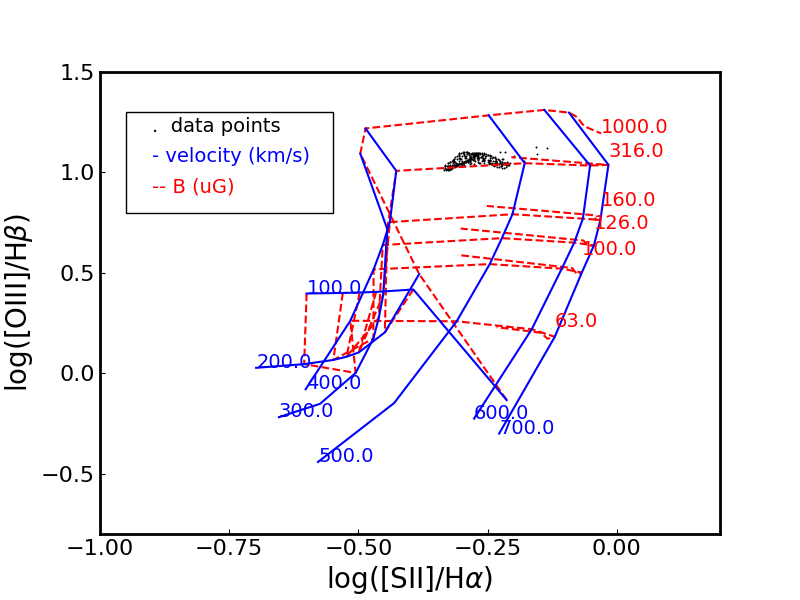}
     }
    \caption{Comparison between predictions of the line ratios due to 
photoionization by AGN (left panel) and shocks  (right panel) and the observed line 
ratios. The clustered black points are the observed data points in central 
1$^{\prime\prime}$$\times$1$^{\prime\prime}$. The model grids are shown for different shock velocities, magnetic
fields and ionization parameters. The typical error of the data points in these plots is 0.01 dex in both axes.}
    \label{figure-13}
\end{figure*}

Similarly, to generate the emission line spectra from shocked material, we used the {\tt MAPPINGS-III} code
again implemented in {\tt ITERA}. We considered shock velocities (v) between 100 and 1000 km s$^{-1}$.
The metallicity was assumed as solar consistent with that available in the literature \citep{2002A&A...391..809C}, 
and we
considered both pure shock and shock plus precursor models. The magnetic parameter B was allowed to
vary between 0.01 to 1000 $\mu$G. We show in Fig. \ref{figure-13}, the comparison between model
line ratios and observed line ratios in the log([OIII]$\lambda$5007/H$\beta$) and
log([SII]/H$\alpha$) plane for photoionization by AGN (left panel) and
photoionization by shocks  (right panel).  The observed line ratios of the spaxels
in the central 1$^{\prime\prime}$$\times$1$^{\prime\prime}$ tend to lie in the region predicted by
shock models. Thus, the observations analysed in this work show evidence of
shocks  contributing to the ionization of the gas in the central region of NGC~4395.
This is possible with the hypothesis that the expanding radio jets from the central core, on its interaction with the ISM, leads to the shocks  in the medium, which dominate the ionization of the gas over other processes, such as photoionization by AGN.

\subsubsection{Electron temperature distribution}
\begin{figure}
    \centering
    \includegraphics[scale=0.45]{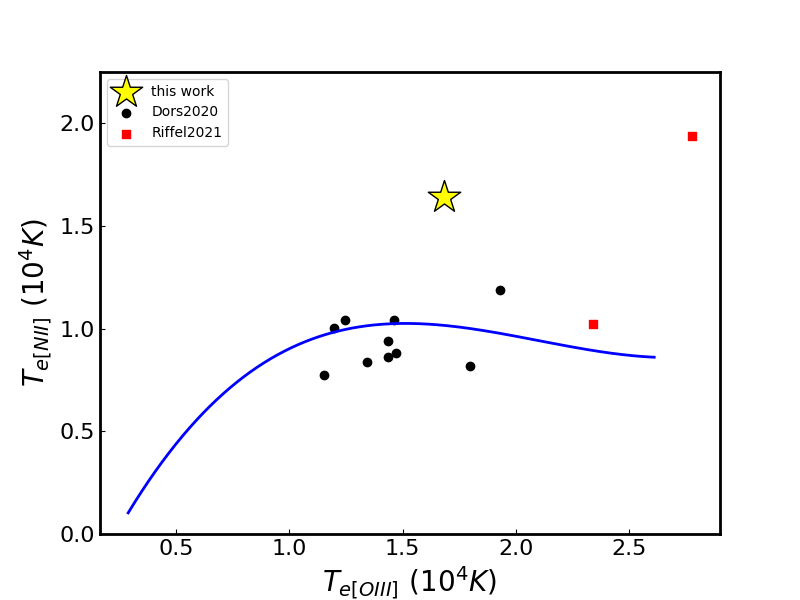}
    \caption{Temperature plot. Here, the yellow point refers to NGC~4395. The blue curve is the AGN photoionization grid. The black and red points are from \cite{10.1093/mnras/staa1781} and \cite{2021MNRAS.501L..54R} respectively.}
    \label{figure-16}
\end{figure}

Knowledge of the electron temperature ($T_e$) in the central regions of AGN can help one to constrain the contribution of AGN to gas ionization.  Shocks from AGN outflows could produce higher values of  $T_e$ \citep{2021MNRAS.501L..54R}. We calculated the integrated $T_e$ using two line intensity ratios namely R$_{O3}$ = ([OIII]$\lambda\lambda$ 4959,5007)/$\lambda$4363 and R$_{N2}$ = ([NII]$\lambda\lambda$6548,6584)/$\lambda$5755 from MaNGA spectra and adopting the following relations (\citealt{2021MNRAS.501L..54R}; \citealt{10.1093/mnras/staa1781}; \citealt{2008MNRAS.383..209H}).
\begin{equation}
\frac{T_{e[OIII]}}{10^4 K} = 0.8254 - 0.0002415R_{O3} + \frac{47.77}{R_{O3}}
\end{equation}
\begin{equation}
\frac{T_{e[NII]}}{10^4 K} = 0.537 + 0.000253 \times R_{N2} + \frac{42.13}{R_{N2}}
\end{equation}
We found $T_{e[NII]}$ = (16.4$\pm$1.3)$\times$10$^3$ K and $T_{e[OIII]}$ = (16.8$\pm$1.0)$\times$10$^3$ K, which are comparable within the errors. 
These values are too large to be produced solely by AGN photoionization (cf. Fig. \ref{figure-16}).

\begin{figure*}
    \centering
    \hspace{-1.0cm}\includegraphics[scale=0.4]{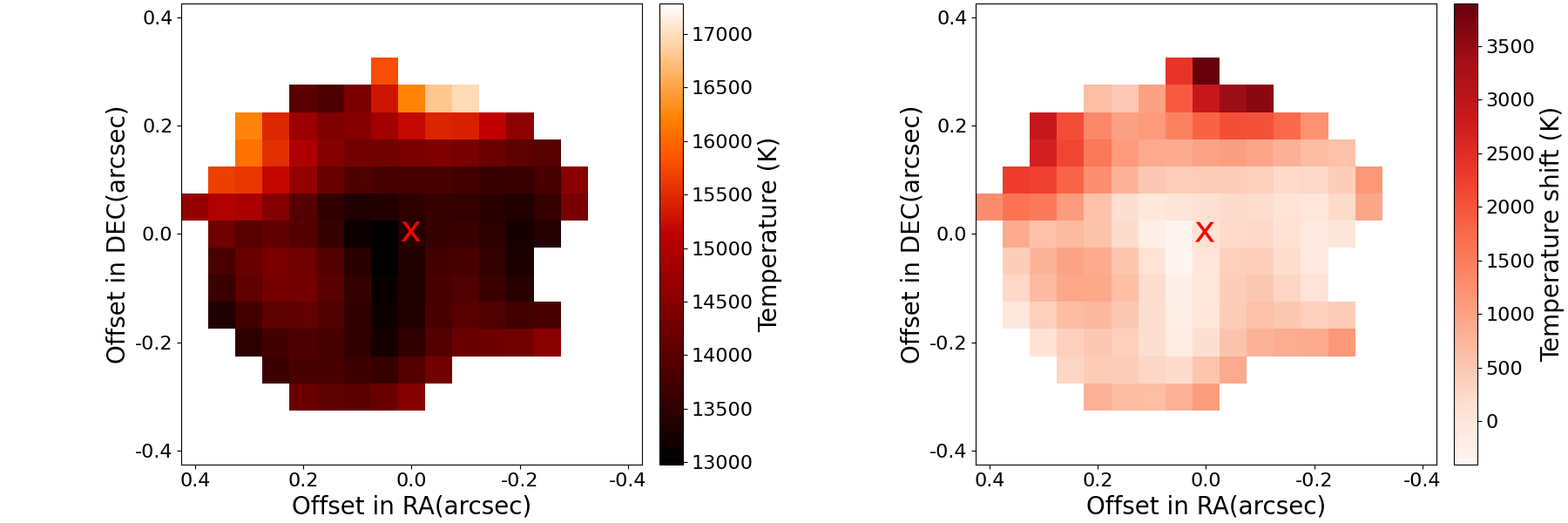}
    \caption{Left panel: Temperature map from [NII] line ratio.  Right panel: 
Temperature difference map with respect to the central spaxel. The red cross is the optical {\it Gaia} position.}
    \label{figure-14}
\end{figure*}

To better characterise the spatial nature of $T_e$, we used [NII] lines from GMOS spectra to generate a spatially resolved map of $T_e$. Since [OIII]$\lambda$4363 is not covered by the GMOS spectra we used the line ratio R$_{N2}$ to generate the $T_e$ map. For this we considered only those spaxels where the S/N ratio (ratio of the peak of the [NII]$\lambda$5755 line to the standard deviation of the pixels in the adjacent continuum) is greater than 30. The $T_e$ map is shown in Fig.~\ref{figure-14}. We found $T_e$ to have a range of values, with the value increasing from the center of NGC~4395 outwards, both towards the eastern and western terminal points of the radio jet. This increase in temperature towards the eastern and western sides is evident in the temperature difference map shown in the right panel of Fig~\ref{figure-14}. This 
temperature difference map is generated by subtracting each temperature value 
from the temperature calculated over the central 0.05$^{\prime\prime}$$\times$0.05$^{\prime\prime}$ region. This difference in temperature is significant, as the error in
the temperature estimated using Equations 3 and 4 is typically around 6\%$-$8\%. 
The increase of temperature from the centre of NGC~4395 towards 
the edges (the difference in temperature is larger
than the error in the estimated temperature) coinciding with the radio jet, 
points to the gas being mostly ionised by shocks. Shocks could be produced by the interaction of the radio jet with the ISM, and this increase of $T_e$ from the centre towards the edges is a direct evidence of shock ionisation \citep{2021MNRAS.501L..54R}.

\subsection{Warm ionized gas and shock} The availability of gas reservoirs in the few tens of parsec in the central regions of AGN is an important ingredient in the feeding and feedback processes in them. In particular, the presence of ionized gas in the central regions of AGN is believed to be a consequence of SF as well as AGN activity. Such ionised emission could also be produced by shock excitation.  The presence of such ionised gas is easily traced in the optical through emission lines and could trace the effect of AGN and the presence of outflows.  
From recent IFU observations in the optical and infrared of the central 
1$^{\prime\prime}$$\times$1$^{\prime\prime}$ region of NGC~4395, 
\citet{2019MNRAS.486..691B} suggest that these may be ionized by the AGN based on the location of these spatially resolved measurements in the BPT diagram \citep{1981PASP...93....5B}
(in the case of optical) and IR line ratio diagram (in the case of infrared). However, in the zoomed in version of the BPT diagram, the eastern component, the core and the western component nicely gets 
segregated (see Fig. \ref{figure-12}). It is thus likely (similar to that seen in a nearby AGN NGC~1068 by \citealt{2019MNRAS.487.4153D}), the emission in the spaxels within the central 1$^{\prime\prime}$$\times$1$^{\prime\prime}$ region of NGC~4395 could have contribution from AGN, besides shock ionisation as discussed earlier.

Outflows can have multiple constituents, such as the hot ionized gas produced at the shock front as well as neutral and molecular gas entrained in the flow. Shocks produced by AGN-driven outflows and/or radio jet-ISM interaction could 
also provide the possibility of energetic feedback altering the SF 
characteristics of the ISM. We consider here the possibility of the shocks 
leading to the observed morphology of the ionized [OIII]$\lambda$5007 

We calculated the mass of the outflowing ionised hydrogen from the measured 
luminosity of the H$\alpha$ emission using the following relation 
\citep{2017A&A...604A.101C}.
\begin{equation}
M_{ion}^{out} = 3.2 \times 10^5\left(\frac{L_{broad}(H\alpha)}{10^{40} erg s^{-1}}\right) \left(\frac{n_e}{100 cm^{-3}}\right)^{-1}
\end{equation}
By considering $F_{broad}({H\alpha}$) = 7.40$\times$10$^{-14}$ erg cm$^{-2}$ s$^{-1}$ (integrated flux density over a circular aperture of radius 0.4$^{\prime\prime}$ 
on the extinction corrected outflowing component of H$\alpha$ line image from 
GMOS), and mean electron density, $n_e$ = 1700 cm$^{-3}$, we obtained 
M$_{ion}$$^{out}$$\sim$ 652M$_{\odot}$. Using a $\sigma$ of 123 km s$^{-1}$ 
(median $\sigma$ of outflowing component of [OIII]$\lambda$5007 line), we 
calculated the kinetic energy of this ionised mass as $E_{KE} = M_{ion}^{out} 
(\sigma^2)$ = 1.97$\times$10$^{50}$ erg. Taking a velocity of 9 km/s (median 
of velocity shift of outflowing component of [OIII]$\lambda$5007 line) and the 
projected distance of the tip of the eastern
jet as 0.3$^{\prime\prime}$ (6.3 parsec), the time required to reach the terminal point 
is 2.16$\times$10$^{13}$ s. The power of the outflow is thus 
$P_{out} = E_{KE}/t$ = 9.14$\times$10$^{36}$ erg s$^{-1}$.

We calculated the mass and radius of the NLR using the following relations 
\citep{1997iagn.book.....P}.
\begin{equation}
M_{NLR} = 7 \times 10^5\left(\frac{L(H\beta)}{10^{41} erg s^{-1}}\right) \left(\frac{10^3 cm^{-3}}{n_e}\right) M_{\odot}
\end{equation}

\begin{equation}
R_{NLR} = 19 \left(\frac{L(H\beta)}{\epsilon 10^{41} erg s^{-1}}\right)^{1/3} \left(\frac{10^3 cm^{-3}}{n_e}\right)^{2/3} parsec 
\end{equation}

By considering $n_e$ of 1700 cm$^{-3}$ (obtained from GMOS observations, 
see Fig.~\ref{figure-6}) and assuming a filling factor ($\epsilon$) of 
10$^{-2}$ (typical upper limit;\citealt{1997iagn.book.....P}) we obtained 
mass and radius of the NLR of NGC~4395 as 282 $M_{\odot}$ and 5.35 parsec 
respectively over a circular region of 0.4$^{\prime\prime}$ radius.

We calculated the bolometric 
luminosity (L$_{Bol}$) using the observed brightness in soft X-ray, hard X-ray 
and H$\alpha$. In the hard X-ray band (14$-$195 keV), using the logarithm of the observed luminosity of 
40.797 \citep{2014ApJ...783..106L}, we obtained L$_{Bol}$ = 
4.97$\times$10$^{41}$ erg s$^{-1}$ using the following relation \citep{2017ApJ...835...74I}:
\begin{equation}
log (L_{Bol}) = 0.0378 \times (log (L_X))^2 - 2.03 \times log (L_X) + 61.6.
\end{equation}

In the soft X-ray band (2$-$10 keV)  using the logarithm of the observed  luminosity of 
40.3 \citep{2011MNRAS.417.2571N}, we obtained a L$_{Bol}$ of 1.95$\times$10$^{41}$ erg s$^{-1}$ using the
relation given below:
\begin{equation}
log(L_{Bol}) = 0.0378 \times log(L_{2-10})^2 - 2.00 \times log(L_{2-10}) + 60.5.
\end{equation}

Similarly, from H$\alpha$ GMOS observations (considering a circular aperture of 0.4$^{\prime\prime}$) using a H$\alpha$  luminosity of
$5.43\times10^{38} erg s^{-1}$ , we obtained L$_{Bol}$ = 3.64$\times$10$^{41}$ erg s$^{-1}$ using the equation given 
below \citep{2007ApJ...670...92G}.
\begin{equation}
L_{Bol} = 2.34 \times 10^{44} \times (L_{H\alpha}/10^{42} erg s^{-1})^{0.86}
\end{equation}

Thus, from optical and X-ray observations,  we found the source to have a
bolometric luminosity in the range of 
(1.95$-$4.97$) \times$10$^{41}$  erg s$^{-1}$.

The disk accretion rate
is generally represented by the Eddington ratio ($\lambda_{Edd}$) and is
defined as

\begin{equation}
\lambda_{Edd} = L_{Bol}/L_{Edd}
\end{equation}
Here, L$_{Edd}$ is the Eddington luminosity defined as

\begin{equation}
L_{Edd} = 1.26 \times 10^{38}\left(\frac{M_{BH}} {M_{\odot}}\right) erg s^{-1}
\end{equation}

Using L$_{Bol}$ of (1.95$-$4.97)$\times$10$^{41}$ erg s$^{-1}$  and M$_{BH}$ 
values of (9.1$\times$10$^3$ - 3.6$\times$10$^5$) M$_{\odot}$ \citep{2019NatAs...3..755W, 2005ApJ...632..799P} we obtained 
$\lambda_{Edd}$ values of 0.004 to 0.044.

Given the jet power and the bolometric luminosity to be larger than the power 
of the outflowing ionized emission, the outflow seen in this source on the 
scale of the NLR of the source could be because of either jet-mode or 
radiative mode process. The optical spectrum from MaNGA for the central region 
encompassing the complete core$-$jet structure having an angular size of 0.5$^{\prime\prime}$, shows the presence of the [OIII]$\lambda$4363 and HeII$\lambda$4686 
lines (see Fig. \ref{figure-8}). The logarithm of the ratio 
between [OIII]$\lambda$4363 and [OIII]$\lambda$5007 lines is $-$1.6; 
HeII$\lambda$4686 and H$\beta$ ratio is $-$0.72 and [OIII]$\lambda$5007 and 
H$\beta$ ratio is 0.86. These line ratios point to the presence of shocks 
(\citealt{2017ApJ...847...41C}; \citealt{2002A&A...383...46M}).% in the central region of NGC~4395.

\begin{figure*}
    \centering
      \hbox{
        \includegraphics[scale=0.42]{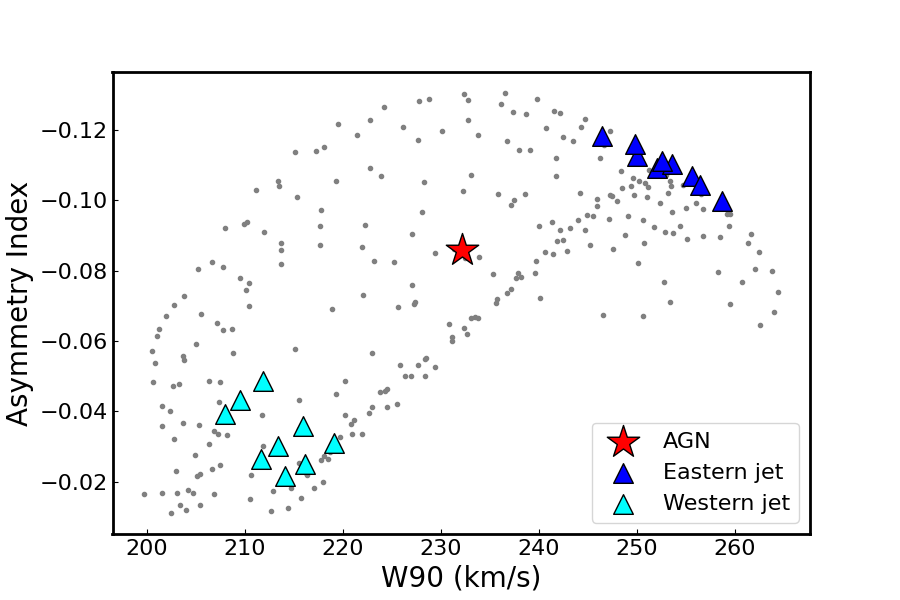}
        \includegraphics[scale=0.42]{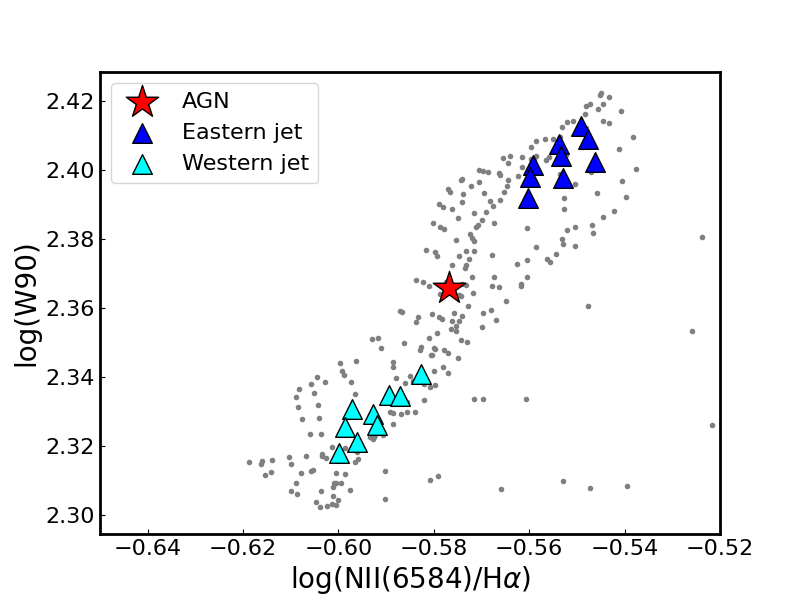}
      }
    \caption{Left Panel: Distribution of spaxels belonging to the 1$^{\prime\prime}$$\times$1$^{\prime\prime}$ 
region of NGC~4395 in the asymmetry index versus the velocity dispersion plane. 
Right Panel: Distribution of spaxels belongings to the 1$^{\prime\prime}$$\times$1$^{\prime\prime}$ 
region in the W90 ([OIII]$\lambda$5007 line) vs. intensity ratio 
of [NII]$\lambda$6584 and H$\alpha$ lines.}
    \label{figure-15}
\end{figure*}

A comparison of the  emission line ratios obtained from photoionization and shock 
modelling and observed line ratios also indicates the gas in the central regions 
of NGC~4395 to be ionised by shocks  (see Fig.~\ref{figure-13}). Assuming a spectral index ($\alpha$) of $-$0.64 \citep{2022MNRAS.514.6215Y} for the eastern jet component we derived the Mach number $\left(M_s= \sqrt{\frac{2\alpha -3}{2\alpha + 1}}\right)$ \citep{2021ApJ...916..102A} of the shock as $M_s$=3.91. In the line ratios diagnostic diagrams, such as the [OIII]$\lambda$5007/H$\beta$ versus [SII]$\lambda$6717,6731/H$\alpha$ and [OIII]$\lambda$5007/H$\beta$ versus [NII]$\lambda$6584/H$\alpha$ diagrams, though all the spaxels lie in the region occupied by AGN, the structure is clearly delineated (see Fig. \ref{figure-12}). Also, in
the asymmetry of the line versus the velocity dispersion diagram (see Fig. \ref{figure-15}; left panel), the spaxels in the eastern jet, occupy a region of higher line asymmetry and
higher velocity dispersion, while the western jet occupies a region of lower asymmetry index and lower velocity dispersion. High velocity dispersion and high asymmetry of the lines are attributed to shock excitation \citep{2019MNRAS.485L..38D}. The eastern
jet thus seems to occupy a region that is dominated by shock excitation, while the western jet seems to occupy a region of weaker shocks.
Spaxels in the central 1$^{\prime\prime}$$\times$1$^{\prime\prime}$ region show a tight correlation between the velocity dispersion and the shock sensitive line ratio [NII]/H$\alpha$ (see Fig. \ref{figure-15}; right panel) (\citealt{10.1093/mnras/stu1653}). Shock models predict an increase in [NII]/H$\alpha$ with an increase in shock velocity \citep{2010A&A...519A..40A}.
We show in Fig. \ref{figure-16} the position of NGC~4395 in the $T_{e[NII]}$ 
versus $T_{e[OIII]}$ diagram estimated from MaNGA spectrum. In the same diagram, 
there are measurements for few AGN along with predictions from AGN photoionization from {\tt MAPPINGS-III}. AGN NGC~4395 lies in a distinct position in 
this Figure, significantly deviant (inclusive of errors in 
the temperature measurement) from the postion occupied by  sources
 photoionized by AGN,  pointing to such high temperatures being produced by shocks.

Photoionization and shock modelling from {\tt MAPPINGS-III} \citep{2013ascl.soft06008S}, the electron temperature distribution and disturbed kinematics point to the gas in the central region of NGC~4935, excited by shocks.
From a multitude of arguments, we conclude that shocks are the dominant process contributing to the excitation of the gas, and such shocks could be due to the interaction of the jet with the ISM in the central 10 parsec region of NGC~4395.

\begin{figure}
\centering
\hspace*{-0.7cm}\includegraphics[scale=0.5]{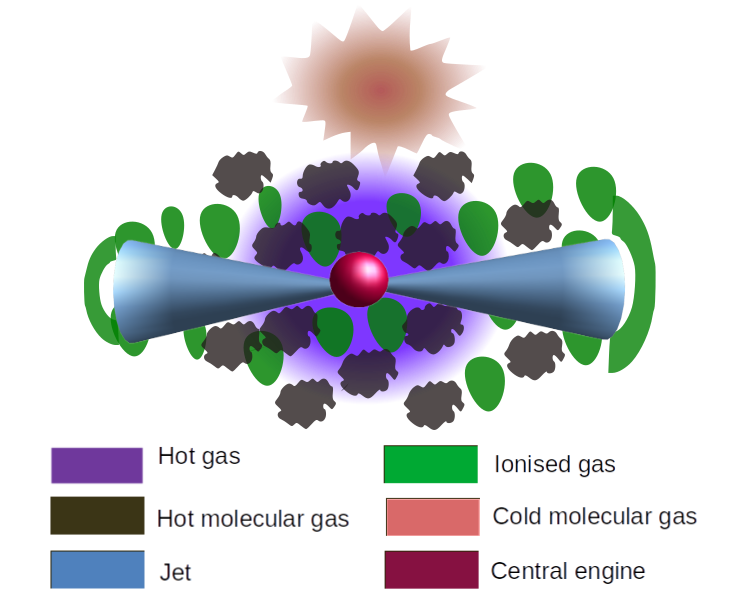}
\caption{Schematic diagram of our proposed scenario in the inner region of NGC~4395. The jet 
on its travels outwards from the central radio core, interacts with the medium and ionizes the
gas via shock excitation. The radio core coincides with the optical {\it Gaia} position, the peak
of the [OIII] emission and the peak of the 237 GHz emission. Ionised [OIII] has a cone-like
structure, with the radio jet along the axis and causing the outflows. The CO(2$-$1) gas is 
located northwards by $\sim$ 1$^{\prime\prime}$ from the radio core. While ionised [OIII]$\lambda$5007, warm molecular H2$\lambda$2.4085 emissions are along the jet, there is a lack of cold CO in the 
vicinity which is possibly due to interactions with the radio jet. As cold gas is needed for SF process,
the lack of cold gas naturally leads to conditions less favourable for SF at scales of 
$\sim$ 10 parsec close to the center of NGC~4395.}
\label{figure-17}
\end{figure}

\subsection{A radio jet-ISM interaction on 10 parsec scale in NGC~4395} 
From an analysis of data in the optical, infrared, radio and sub-mm bands, we have 
found evidence of a low-luminosity jet interacting with its host on the scale  of about 10 parsec. The eastern jet component which is brighter in the radio band, 
is resolved in the high-resolution High Sensitive Array (HSA) image into two 
components oriented approximately in the North-South direction, which is nearly 
orthogonal to the source axis \citep{2006ApJ...646L..95W}. This indicates 
interaction of the jet plasma with the ISM, with the plasma following the path 
of least resistance. On the eastern side, the [OIII]$\lambda$5007 line-emitting 
gas has higher velocity, higher velocity dispersion and higher asymmetry (see 
Fig. \ref{figure-15}, left panel), possibly due to the shocks associated with the interaction of the radio plasma with the [OIII]$\lambda$5007 gas. The weakness of the [OIII]$\lambda$5007 emission can either be due to the gas being more ionized or larger extinction, E(B$-$V) (see Fig. \ref{figure-6}) or a combination of both. The weaker jet on the western side has a smaller effect on the [OIII]$\lambda$5007 gas with a lower velocity dispersion and asymmetry. This suggests that there may be an intrinsic asymmetry in the oppositely-directed jets. The radio emission is found to exist co-spatially with the emission at other wavelengths such as the hot ionised [OIII]$\lambda$5007 emission in the optical
band, the warm molecular H$_2$$\lambda$2.4085 in the infrared band and the 237 GHz emission in the sub-mm band. However, the cold CO(2$-$1) emission is displaced by $\approx$1$^{\prime\prime}$ from the core. The presence of cold molecular gas is conducive for star formation. The displacement of the
CO(2$-$1) gas, and the paucity of cold gas along the source axis, possibly due to interactions by the jet, has led to the conditions less favourable
for SF on 10 parsec scale. A schematic of our proposed coherent picture of the central region of NGC~4395 is shown in Fig. \ref{figure-17}.

\begin{figure*}
   \vspace*{-0.2cm}
\hspace*{1.5cm}  \includegraphics[scale=0.7]{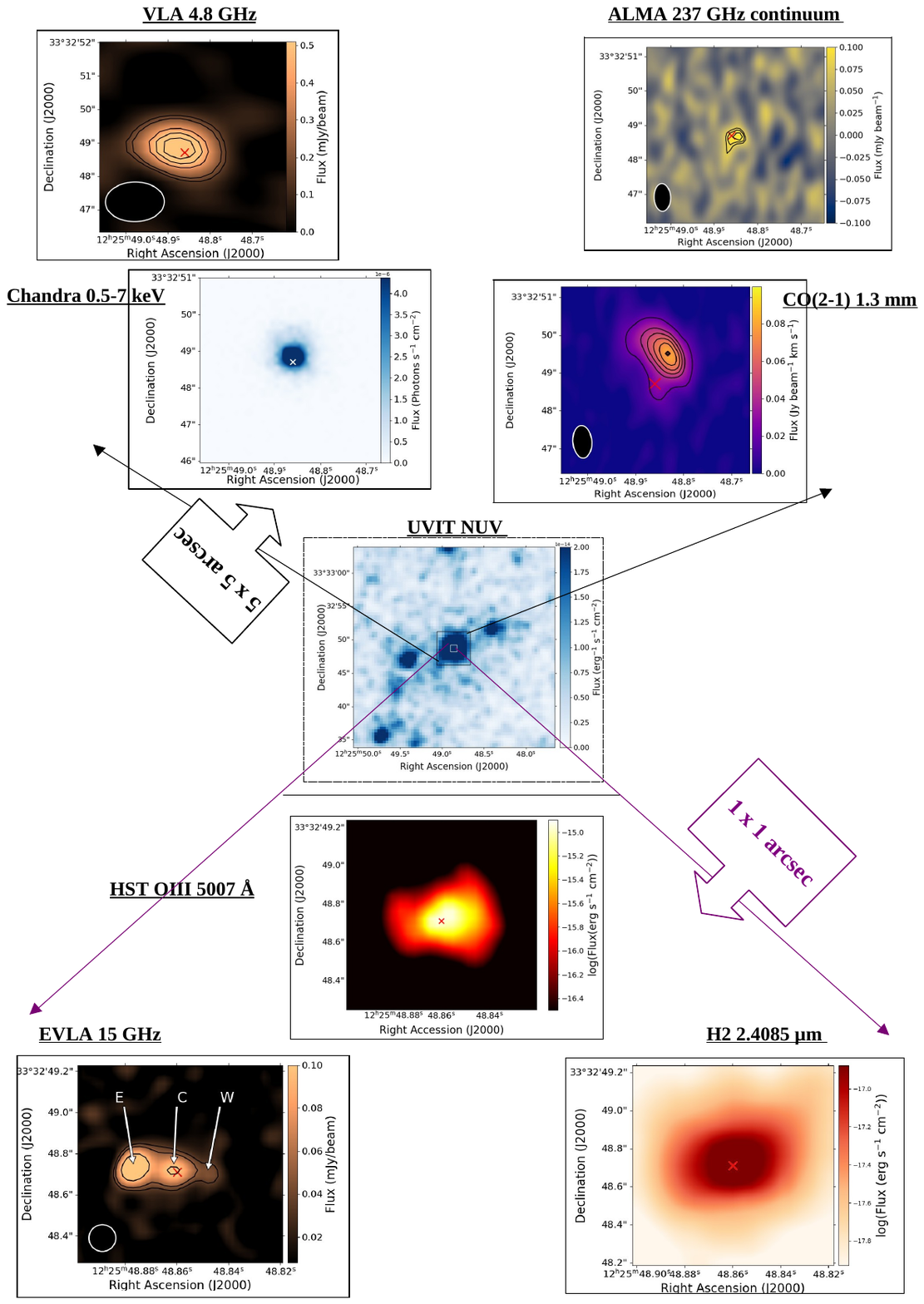}
    \vspace*{1.5cm}
\caption{The comprehensive picture of the jet$-$ISM interaction in the central $\sim$ 10 parsec
region of NGC~4395.  In the center is the image of NGC~4395 in the NUV filter 
observed with UVIT \citep{2023ApJ...950...81N}. This image has a size of 30$^{\prime\prime}$$\times$30$^{\prime\prime}$. Also, shown in the same image are two square boxes one of
size 5$^{\prime\prime}$$\times$5$^{\prime\prime}$ (black colour) and the other of 1$^{\prime\prime}$$\times$1$^{\prime\prime}$ (white colour).
On the top panels are the 4.8 GHz image from the VLA, 237 GHz image
from ALMA, the CO(2-1) image from ALMA and the X-ray image in the 0.5$-$7 keV from
{\it Chandra}. These images have a size of 5$^{\prime\prime}$$\times$5$^{\prime\prime}$. The bottom panels show the
15 GHz image from the VLA, molecular H2$\lambda$2.4085 image from
Gemini and [OIII]$\lambda$5007 image from HST 
over a 1$^{\prime\prime}$$\times$1$^{\prime\prime}$ region. In each of these images, the core of 
NGC~4395 taken as the \textit{Gaia} position, is shown as a cross.}
    \label{figure-18}
\end{figure*}

\section{Summary}
In this work we carried out a systematic investigation of the central
region of NGC~4395 using imaging and spatially resolved spectroscopic
observations. We summarize our main findings below:
\begin{enumerate}
\item From VLA images at 15 GHz, NGC~4395 is found to show a triple radio
structure having a projected size of $\sim$ 10 parsec. The central component
of the triple structure is found to coincide with the optical 
{\it Gaia}
position which we call as the radio core. The source is also highly asymmetric
in brightness with the eastern component brighter than the western one.

\item The triple radio structure in NGC~4395 is reminiscent of bipolar
jet ejection in radio-loud AGN and the eastern and western components of
this triple structure are formed by the low power jet 
(P$_{jet}$ = (1.3 $\pm$ 0.3)$\times$10$^{40}$ erg s$^{-1}$) powered by the intermediate-mass black hole in NGC~4395.

\item From HST observations we found the [OIII]$\lambda$5007 emission to be 
prevalent over the entire extent of the radio emission with the peak of the 
[OIII]$\lambda$5007 emission
coinciding with the optical {\it Gaia} position and the radio core.
The [OIII]$\lambda$5007 emission is asymmetric and shows a convex-shaped structure at the
terminal points on either side of the core of NGC~4395 indicating an outflow. 
This asymmetry in the brightness of [OIII]$\lambda$5007 emission could be due to 
intrinsic asymmetries causing different levels of ionisation or differences in the degree of extinction or a combination of both. 

\item The X-ray emission in the 0.5$-$7 keV band is found to overlap with the radio jet. Similarly, the peak of the continuum emission at 237 GHz is spatially coincident with the radio core. Also, the molecular H$_2$$\lambda$2.4085 is found 
to be extended along the radio jet direction and have a close correspondence with the radio emission.

\item From AGN photoionization and shock modelling from {\tt MAPPINGS-III} and the distribution of the electron temperature distribution, 
we conclude that the gas in the central region of NGC~4395 is excited mostly by 
shocks and such shocks could be due to the interaction of radio jet with the 
ISM in the central parsec region of NGC~4395. This is the first detection of 
radio jet - ISM interaction at such small spatial scales.

\item The cold CO(2$-$1) emission is found to be displaced northwards of the
radio core by about 1$^{\prime\prime}$ ($\sim$ 20 pc). 
The paucity of cold molecular gas in the central region, possibly due to interactions with the jet makes conditions less favourable for star formation
on scales of about 10 parsec in NGC~4395.
\end{enumerate}

The detections of AGN and intermediate mass black holes in a number of dwarf galaxies in recent years have opened the possibility of studying feedback processes in dwarf galaxies. Studies of nearby dwarfs also enable us to probe feedback processes on parsec scales. \citet{2022Natur.601..329S} reported a 150 parsec long ionized filament in the dwarf galaxy Henize 2-10 from HST observations, which connect the black hole region with a site of recent SF. \citet{2017ApJ...845...50N} reported possible evidence of shock excitation in the nearby dwarf AGN galaxy NGC~404 with an amorphous radio outflow extending over $\sim$ 17 parsec. NGC~4395 is the clearest example of a dwarf AGN with a triple radio structure, where there is clear evidence of 
jet$-$ISM interaction on the smallest scale of $\sim$ 5 parsec on either side of the core. A comprehensive picture depicting this jet$-$ISM interaction
in the central parsec scale region based on the analysis of images from 
multiple wavelengths is shown in Fig. \ref{figure-18}.
This finding will bolster the prospect of finding more such instances in dwarf AGN host galaxies,  paving the way for a better understanding of the complex interplay between AGN and their hosts on such small scales in these galaxies.

\vskip12pt
\newpage

\section*{Acknowledgments}
We acknowledge the reviewer for his/her insightful comments, which helped to improve the quality of the manuscript. This research has made use of data obtained from the Chandra Data Archive and the Chandra Source Catalog and software provided by the Chandra X-ray Center (CXC) in the application packages CIAO and Sherpa. This research is based on observations made with the NASA/ESA Hubble Space Telescope obtained from the Space Telescope Science Institute, which is operated by the Association of Universities for Research in Astronomy, Inc., under NASA contract NAS 5$–$26555. This work is partly based on observations obtained at the Gemini Observatory, which is operated by the Association of Universities for Research in Astronomy, Inc., under a cooperative agreement with the NSF on behalf of the Gemini partnership: the National Science Foundation (United States), the Science and Technology Facilities Council (United Kingdom), the National Research Council (Canada), CONICYT (Chile), the Australian Research Council (Australia), Ministério da Ciênciae Tecnologia (Brazil) and south-east CYT (Argentina). 
This publication uses data from MaNGA (Mapping Nearny Galaxies at APO) survey which is one of the programe of Sloan Ditial Sky Survey (SDSS) IV. Funding for the Sloan Digital Sky Survey IV has been provided by the 
Alfred P. Sloan Foundation, the U.S. Department of Energy Office of Science, and the Participating Institutions. 

SDSS-IV acknowledges support and resources from the Center for High Performance Computing  at the University of Utah. The SDSS website is www.sdss4.org.

SDSS-IV is managed by the Astrophysical Research Consortium for the Participating Institutions of the SDSS Collaboration including the Brazilian Participation Group, the Carnegie Institution for Science, Carnegie Mellon University, Center for Astrophysics | Harvard \& Smithsonian, the Chilean Participation Group, the French Participation Group, Instituto de Astrof\'isica de Canarias, The Johns Hopkins University, Kavli Institute for the Physics and Mathematics of the Universe (IPMU) / University of Tokyo, the Korean Participation Group, Lawrence Berkeley National Laboratory, Leibniz Institut f\"ur Astrophysik Potsdam (AIP),  Max-Planck-Institut f\"ur Astronomie (MPIA Heidelberg), Max-Planck-Institut f\"ur Astrophysik (MPA Garching), Max-Planck-Institut f\"ur Extraterrestrische Physik (MPE), National Astronomical Observatories of China, New Mexico State University, New York University, University of Notre Dame, Observat\'ario Nacional / MCTI, The Ohio State University, Pennsylvania State University, Shanghai Astronomical Observatory, United Kingdom Participation Group, Universidad Nacional Aut\'onoma de M\'exico, University of Arizona, University of Colorado Boulder, University of Oxford, University of Portsmouth, University of Utah, University of Virginia, University of Washington, University of Wisconsin, Vanderbilt University, and Yale University.

This paper makes use of the following ALMA data: ADS/JAO.ALMA\#2017.1.00572.S. ALMA is a partnership of ESO (representing its member states), NSF (USA) and NINS (Japan), together with NRC (Canada), MOST and ASIAA (Taiwan), and KASI (Republic of Korea), in cooperation with the Republic of Chile. The Joint ALMA Observatory is operated by ESO, AUI/NRAO and NAOJ. This publication uses radio observations carries out using the National Radio Astronomy Observatory facilities Very Large Array (VLA) and Karl G. Jansky Very Large Array (JVLA). The National Radio Astronomy Observatory is a facility of the National Science Foundation operated under cooperative agreement by Associated Univerties, Inc.This work has made use of the NASA Astrophysics Data System (ADS)\footnote{https://ui.adsabs.harvard.edu/} and the NASA/IPAC extragalactic database (NED)\footnote{https://ned.ipac.caltech.edu}. 
This work has made use of data from the European Space Agency (ESA) mission Gaia (https://www.cosmos.esa.int/gaia), processed by the Gaia Data Processing and Analysis Consortium (DPAC, https://www.cosmos.esa.int/web/gaia/dpac/consortium). Funding for the DPAC has been provided by national institutions, in particular, the institutions participating in the Gaia Multilateral Agreement. The authors acknowledge Dr. Jarle Brinchmann for sharing the library of {\tt ITERA} code. A few of the authors thank the Alexander von Humboldt Foundation, Germany, for the award of the Group Linkage long-term research program. PN thanks the Council of Scientific and Industrial Research (CSIR), Government of India, for supporting her research under the CSIR Junior/Senior research fellowship program through the grant no. $09/079(2867)/2021-EMR-I$.

\software{IRAF \citep{1986SPIE..627..733T}, Astropy \citep{2013A&A...558A..33A,2018AJ....156..123A, 2022ApJ...935..167A}, 
Scipy \citep{2020SciPy-NMeth}, Numpy \citep{harris2020array}, Matplotlib \citep{Hunter:2007}, PyNeb \citep{2015A&A...573A..42L},
AIPS \citep{1985daa..conf..195W}, CASA \citep{2007ASPC..376..127M, 2022PASP..134k4501C}, 
Chandra \citep{2006SPIE.6270E..1VF}, MAPPINGS-III \citep{2013ascl.soft06008S}.
}

\bibliography{ref1}{}
\bibliographystyle{aasjournal}
\end{document}